\documentclass[%
 preprint,
 amsmath,amssymb,
 aps,
]{revtex4-2}

\usepackage{graphicx}
\usepackage{dcolumn}
\usepackage{bm}
\usepackage[dvipsnames]{xcolor}
\usepackage{hyperref}
\usepackage{physics}
\usepackage{slashed}

\usepackage{mathtools}
\usepackage[export]{adjustbox}

\begin{document}

\preprint{UCI-HEP-TR-2023-12}

\title{Conserved Currents are Not Anomaly-Safe}

\author{Tyler B. Smith}
\email{tylerbs@uci.edu}
\affiliation{Department of Physics and Astronomy, University of California, Irvine, CA 92697, USA}

\author{Tim M.P. Tait}%
\email{ttait@uci.edu}
\affiliation{Department of Physics and Astronomy, University of California, Irvine, CA 92697, USA}

\date{\today}

\begin{abstract}
New vector bosons that are coupled to conserved currents in the Standard Model exhibit enhanced rates below the electroweak scale from anomalous triangle amplitudes, leading to (energy/vector mass)$^2$ enhancements to rare Z decays and flavor-changing meson decays into the longitudinally polarized vector boson.  In the case of a vector boson gauging $U(1)_{B-L}$, the mass gap between the top quark and the remaining SM fermions leads to (energy/vector mass)$^2$ enhancements for processes with momentum transfer below the top mass.  In addition, we examine the case of an intergenerational $U(1)_{B_3 - L_2}$ that has been proposed to resolve the $(g-2)_\mu$ anomaly with an MeV scale DM candidate, and we find that these enhanced processes constrain the entire parameter space.
\end{abstract}

\maketitle

\newpage
\section{Introduction}

The standard model (SM) of particle physics, while a monumental triumph describing a vast range of phenomena over many scales, remains incomplete. Notable gaps range from the mysterious nature of dark matter (DM) and dark energy, to the mechanism behind baryogenesis, and the unresolved origins of neutrino masses. A very common element among proposed solutions to these issues is the introduction of new vector bosons (generically denoted as $X$) corresponding to a new $U(1)_X$ gauge symmetry, whose masses can be engineered to lie over a wide range. An important theoretical constraint on the construction of such theories is the need to design their field content such that they are mathematically self-consistent.  For that reason,
new vector bosons coupled to conserved currents within the Standard Model are particularly compelling because their introduction does not require additional fermions for a consistent theory. 

For example, while both baryon- and lepton-number are global symmetries of the SM at tree-level, they cannot be consistently gauged at the quantum level. However, the linear combination $B-L$ is anomaly-free with regards to the SM, though the $U(1)^3_{B-L}$ cubed anomaly necessitates the introduction of right-handed (RH) neutrinos to cancel its internal anomalies.  Since RH neutrinos are a very typical ingredient of theories generating masses for the SM neutrinos, the need to introduce them could be seen as further strengthening the case for extending the Standard Model by gauging $B-L$. 

Of course, a current that is not naively conserved within the Standard Model can nonetheless be gauged by incorporating new fermions (sometimes referred to as ``anomalons") to cancel the gauge anomalies.  This results in a theory in which the current under consideration can be self-consistently gauged, but opens the door to important constraints when these new fermions are massive enough such that there is a mass gap between them and the SM fermions \cite{Dror2017a,Dror2017b}.  At energies far below the mass of the anomalons, their finite contributions to triangle diagrams effectively decouple, leading the current to be effectively non-conserved at the quantum level.  The non-zero divergence of the current in turn leads to processes mediated by the same triangle diagram encapsulating the anomaly experiencing an enhancement of $( E_X / m_X)^2$ for the longitudinal $X$ polarization, where $E_X$ and $m_X$ are the energy and mass of the new $X$ gauge boson. 

In this work, we extend this logic to the conserved currents of the Standard Model itself: the top quark, whose mass is $\sim 40$ times that of the next heaviest fermion in the SM, the bottom quark, contributes a similar decoupling finite contribution to the triangle diagram, leading to a similar enhancement for processes taking place at energies well below the top mass, but significantly above $m_X$.  We perform a thorough analysis of the implications of this phenomenon for the explicit case of $U(1)_{B-L}$, though it is important to note that this is a broader phenomena that is applicable to other conserved currents within the SM.

The paper is structured as follows, in Section \ref{sec:anomalousamps} we provide a short review of the calculation of the triangle diagrams contributing to gauge anomalies, illustrating the decoupling phenomenon for heavy fermions. In Section \ref{sec:processes} we discuss the two enhanced processes which primarily constrain the parameter space of $U(1)_{B-L}$. The impact of the experimental constraints on $U(1)_{B-L}$ are detailed in Section \ref{sec:experiments}. We offer an additional example of an inter-generational $B_i-L_j$ application in \ref{sec:b3-l2}. We conclude with summary and outlook in Section \ref{sec:discussion}.

\section{Anomalous Amplitudes}
\label{sec:anomalousamps}

\begin{figure}[th]
  \centering\small
  \includegraphics[width=7cm, valign=c]{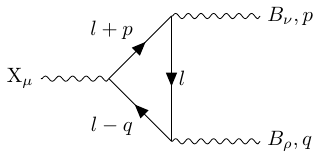}
  \qquad 
  $+$
  \qquad 
  \includegraphics[width=7cm, valign=c]{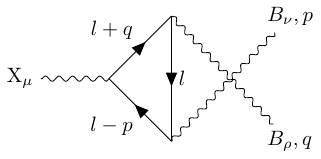}
  \caption{The anomalous triangle diagrams for the case of $U(1)_Y^2$-$U(1)_X$, where B denotes the hypercharge boson. Corresponding diagrams hold for the case of the hypercharge bosons being substituted by the $SU(2)_L$ gauge bosons.}
  \label{fig:XBBfeynmandiagram}
\end{figure}

We are interested in theories which gauge a global symmetry of the Standard Model under which the Standard Model fermions are vector-like.  Such a
symmetry may still potentially have mixed anomalies with the Standard Model's $SU(2)_L \times U(1)_Y$ gauge symmetries because of the chiral SM electroweak interactions.  In the UV, where the effect of electroweak symmetry breaking can be neglected, the statement that the theory is free from $SU(2)_L^2$- $U(1)_X$ and $U(1)_Y^2$-$U(1)_X$ mixed anomalies implies that the charges of the fermions under $U(1)_X$ are such that the triangle diagrams (Fig \ref{fig:XBBfeynmandiagram}) vanish after summing across all of the standard model fermions.  However, below the scale of electroweak symmetry-breaking, the masses of the fermions running in the loop become relevant. In particular, at energies far below the mass of the top quark, its contribution effectively decouples, effectively leaving behind a non-zero triangle amplitude.

For processes at energies well above the $X$ mass, the Goldstone equivalence theorem \cite{Cornwall:1974km} implies that the longitudinal polarization of $X$ dominates over the transverse modes, and can be well-approximated by its corresponding Goldstone boson via $g_X X_\mu \rightarrow \partial_\mu \varphi / f_X$, where $f_X = m_X /g_X$.  Integrating by parts results in an effective vertex of the form of $\varphi$ multiplied by the divergence of the current to which $X$ couples.  Thus, the regime in which triangle amplitudes are non-vanishing can result in particularly important contributions.

Details of the calculation of the triangle amplitudes including the dependency on the fermion masses can be found in Appendix~\ref{app:anomalyamps}.  The effective vertices for the triangle processes involving $U(1)_Y^2-U(1)_X$ and $SU(2)_L^2-U(1)_X$ are:
\begin{eqnarray}
    (p+q)_\mu \mathcal{M}^{\mu \nu \rho}_{XBB} & = & \sum_f \frac{i g_X g_V g_A }{\pi^2} X 
    ~ \left[ 1 - 2m_f^2 I_{00}(p, q, m_f) \right] 
    ~ \epsilon^{\alpha \beta \nu \rho} q_\alpha p_\beta ,
\label{eq:longampXBB} \\
    (p+q)_\mu \mathcal{M}^{\mu \nu \rho}_{XWW} & = & -\frac{1}{4}\sum_f \frac{i g_X g_2^2}{\pi^2}  \Tr [(X)T^aT^b]
    ~ \left[ 1 - 2m_f^2 I_{00}(p, q, m_f) \right]
    ~ \epsilon^{\alpha \beta \nu \rho} q_\alpha p_\beta
\label{eq:longampXWW}
\end{eqnarray}
with
\begin{equation}
    I_{00}(p, q, m_f) \equiv \int^1_0 dx \int^{1-x}_0 \frac{dz}{m_f^2 - z(1-z)p^2 - x(1-x)q^2 - 2xz (p \cdot q)} .
\end{equation}
Both vertices are proportional to $(1-2m_f^2I_{00})$. The first term is independent of the fermion mass, and sums (over all SM fermions) to zero since the mixed $U(1)_Y^2-U(1)_X$ and $SU(2)_L^2-U(1)_X$ anomalies are absent. The $-2m_f^2I_{00}$ term has two important limits depending on the mass of the fermion in the loop compared to the external momenta. 
In the the massless fermion limit, $m_f^2 \ll p^2,q^2$, $-2m_f^2I_{00} \rightarrow 0$, whereas for fermion masses much greater than the external momenta, $m_f^2 \gg p^2,q^2$, $-2m_f^2I_{00} \rightarrow -1$, which cancels the mass-independent term and implements the expected decoupling of the ultra-heavy fermion.  Putting this together, in the deep UV or IR, the summed contributions of all of the SM fermions will lead to vanishing amplitudes, and no enhanced vertices.  In the intermediate momentum regime where some of the SM fermions are decoupled whereas others are active, there can be enhanced contributions to the effective vertices.

Leveraging the mass of the top quark, which is $\sim 40$ times greater than that of the bottom, we anticipate that there are likely to be important
constraints for processes whose momenta lie within the gap between the two.  These energies have been extensively explored experimentally, and rare $Z$ and flavor-changing meson decays can be particularly constraining. 

\section{Enhanced X Boson Production Processes}
\label{sec:processes}

In this section, we review the rates of specific processes producing $X$ bosons via $Z$ boson and flavor-changing meson decays.

\subsection{ \texorpdfstring{$Z \rightarrow X \gamma$}{Lg} }

In the mass basis, the amplitudes Eqs.~(\ref{eq:longampXBB}) and (\ref{eq:longampXWW}) translate into a vertex contributing to the decay $Z \rightarrow X \gamma$:
\begin{equation}
    (p+q)_\mu \mathcal{M}^{\mu \nu \rho}_{ZX\gamma}= \sum_f \frac{g_X e^2 X}{ \pi^2 S_w C_w} \epsilon^{\alpha \beta \rho \nu} q_\alpha p_\beta (1-2m_f^2 I_{00}) 
     \Bigg\{\sum_{RH}(S_w^2 Y_R^2) - \sum_{LH} (S^2_w Y^2_L - \frac{1}{8} C_w^2) \Bigg\} .
\end{equation} 
Since $m_t \gtrsim m_Z \gtrsim m_b$, the amplitude is well-approximated by considering the top as decoupled and the remaining SM fermions as massless. 
In that limit, the branching ratio for the rare decay $Z \rightarrow X \gamma$ is given by:
\begin{equation}
    \text{BR}(Z \rightarrow X \gamma) \simeq 4 \times 10^{-10} 
    \left[ 1- \left(\frac{m_X}{m_Z}\right)^2 \right]^3 
    \left( \frac{\text{TeV}}{ f_X} \right)^2 .
    \label{eq:Zbranch}
\end{equation}

\subsection{Meson Decays and the Flavor Changing Neutral Current}

\begin{figure}
    \centering
    \includegraphics[width=\columnwidth]{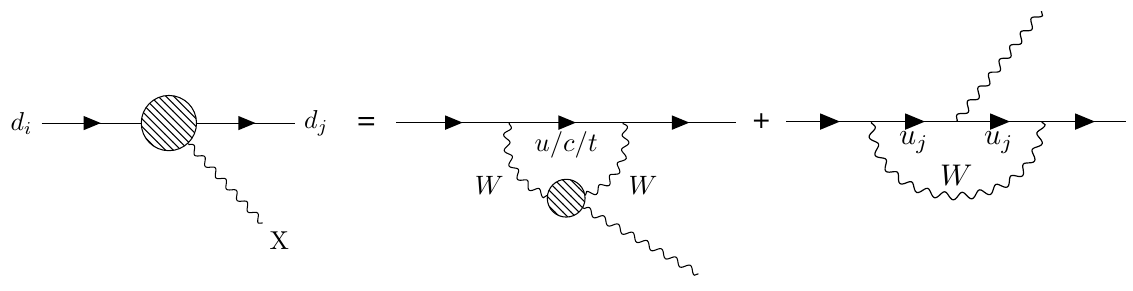}
    \caption{The effective vertex for meson decays. If the current of the X boson is coupled directly to quarks then the second diagram in the summation is present, if not then the only contribution comes from the effective vertex in the first diagram in the summation.}
    \label{fig:FCNC}
\end{figure}

The amplitude for $X \rightarrow WW$ given by Eq.~(\ref{eq:longampXWW}) can potentially give rise to flavor changing neutral currents (FCNCs) via the two diagrams shown in Fig.~\ref{fig:FCNC}~\cite{Dror2017b}.  It is noteworthy that even in the case where $U(1)_{X}$ has flavor-diagonal couplings to the quarks, there are still FCNC contributions resulting from the flavor violation in the SM charged current interactions of the $W$ with the quarks.  In fact, even when the SM quarks decouple entirely from $X$, there will nonetheless be contributions from the first diagram of Fig.~\ref{fig:FCNC}~\cite{Dror2017b}, where the blob stands in for the triangle vertices that are our focus.

The amplitude, to leading order in an expansion $1/m_X$ and neglecting the small masses of the down-type quarks, is:
\begin{equation}
    \mathcal{M} = -i\sum_\alpha^{u,c,t} \frac{V_{\alpha j} V^*_{\alpha i} g^4_2}{32 f_X} \Bar{u}(p_2) \gamma^{\beta} (p_{1,\beta}-p_{2,\beta}) P_L u(p_1) f(x) ,
    \label{eq:ampred}
\end{equation}
where:
\begin{equation}
    f(x) \equiv \frac{x (1 + x(\text{log}(x)-1))}{(1-x)^2} ,
    \label{eq:PVfunc}
\end{equation}
with the variable x given by $x \equiv m_\alpha / m_W$. 

These amplitudes can be described by the effective Lagrangian:
\begin{equation}
    \mathcal{L} \supset g_{X d_i d_j} X_\mu \Bar{d}_j \gamma^\mu \mathcal{P}_L d_i + h.c. .
    \label{eq:effectivelagrangianFCNC2}
\end{equation}
where each coupling $g_{X d_i d_j}$ is obtained by matching to Eq.~(\ref{eq:ampred}).  These effective quark-level couplings can be translated
into decay widths for flavor-changing meson decays making use of QCD light-cone sum rules \cite{Ball:2004rg,Ball:2004ye}, resulting in decay widths:

\begin{eqnarray}
\label{eq:mesonwidths}
\Gamma (B \rightarrow K X) & \simeq & \frac{m_B^3}{64 \pi m_X^2} |g_{bsX}|^2 \left( 1 - \frac{m_K^2}{m_B^2}\right)^2 | f_K(m_X^2)|^2 \frac{2 Q}{m_B} , \label{eq:BtoKX}\\
\Gamma (B \rightarrow K^* X) & \simeq & \frac{m_B^3}{64 \pi m_X^2} |g_{bsX}|^2 | f_{K^*}(m_X^2)|^2 \left(\frac{2 Q}{m_B}\right)^3 , \label{eq:BtoKstarX}\\
\Gamma (K^{\pm} \rightarrow \pi^{\pm} X) & \simeq & \frac{m_{K^\pm}^3}{64 \pi m_X^2} |g_{sdX}|^2 \left( 1 - \frac{m_{\pi^{\pm}}}{m_{K^{\pm}}^2}\right)^2 \frac{2 Q}{m_{K^{\pm}}} , \label{eq:KtopiX}\\
\Gamma (K_L \rightarrow \pi^{0} X) & \simeq & \frac{m_{K_L}^3}{64 \pi m_X^2} \text{Im}(g_{sdX})^2 \left( 1 - \frac{m_{\pi^{0}}}{m_{K_{L}}^2}\right)^2 \frac{2 Q}{m_{K_L}} ,\label{eq:KLtopi0X}
\end{eqnarray}
where $Q$ is the magnitude of the momentum of the decay products in the center-of-mass frame. The form factors $f_{K,K^*}(m_X^2)$ represent hadronic matrix elements found in \cite{Ball:2004rg,Ball:2004ye}.  Some details of their application to the particular case at hand are shown in Appendix~\ref{app:formfacs}. 

\section{Experimental Constraints}
\label{sec:experiments}

In this section we derive specific experimental constraints on  $U(1)_{B-L}$ gauge bosons. Note that the $U(1)_{B-L}$ gauge boson couples to every SM fermion and thus is constrained even in the absence of the introduction of new heavy fermions, dark states, or kinetic mixing (which we assume to be negligibly small) with hypercharge. We discuss the specific constraints from each channel below, and a summary of their impact on the parameter space of $U(1)_{B-L}$ is shown in Fig.~\ref{fig:reachplot}. 

\subsection{Z Decays}

When $X$ decays invisibly (for example, promptly into SM neutrinos, or with a long decay length beyond the detectors), it contributes to the rare decay $Z \rightarrow \gamma + \text{inv}$.
The most stringent existing constraints are from LEP I measurements at the $Z$ pole. In particular, the L3 detector searched for mono-photons with $E_\gamma > 15$ GeV and derive limits on the branching ratio: BR$(Z \rightarrow \gamma + \text{inv}) \lesssim 10^{-6}$ \cite{L3:1997exg} at the $95\%$ confidence limit.

Visible decays of the $B-L$ boson via $Z$ boson hadronic decays are also constrained by LEP. Specifically, the L3 detector explores the $Z$ boson's hadronic decays by searching for single photons with $E_\gamma > 5$ GeV alongside a jet \cite{Rind:1992es}. They place an upper limit on the branching ratio $\lesssim 3.5 \times 10^{-3}$.

Leptonic $Z$ decays are probed by the OPAL detector at LEP in a search for a new gauge boson $X$ producing a resonance in $\ell^+ \ell^-$, where $\ell = e,\mu$. The constraints they place are BR$(Z \rightarrow \gamma (X \rightarrow \ell^+ \ell^-)) \lesssim 1.1 \times 10^{-4}$ in the mass region $60 < m_{X}/\text{GeV} < 82$ \cite{OPAL:1991acn}. The L3 detector expands on this in the energy range of $30 < m_{X}/\text{GeV} < 89$ with slightly weaker constraints of BR$(Z \rightarrow \gamma (X \rightarrow e^+ e^-)) < 2.8 \times 10^{-4}$ and BR$(Z \rightarrow \gamma (X \rightarrow \mu^+ \mu^-)) < 2.3 \times 10^{-4}$ \cite{L3:1991kow}. Additionally, the branching ratio to hadrons has a slight improvement from L3 yielding BR$(Z \rightarrow \gamma (X \rightarrow \text{hadrons})) < 4.7 \times 10^{-4}$ in the mass range $30 < m_{X}/\text{GeV} < 86$. 

\subsection{Meson Decays}

$X$ bosons decaying both into visible and invisible channels can be produced in a variety of meson decays.  We find that decays of kaons and $B$ mesons provide important constraints on $X$ bosons coupled to $B-L$ over wide regions of parameter space.

Invisible channels stemming from $B$ meson decays can be probed at Belle which places $95\%$ CL constraints on the $B$ meson branching ratios: BR$(B \rightarrow K (X\rightarrow \text{inv})) <1.6 \times 10^{-5}$  and BR($B \rightarrow K^* (X\rightarrow \text{inv}) < 2.7 \times 10^{-5}$ \cite{Belle:2017oht}. Belle also places constraints on the visible decays, particularly: BR$(B^+ \rightarrow K^+ (X \rightarrow l^+ l^-))  = (5.99^{+0.45}_{-0.43} \pm 0.14) \times 10^{-7}$ and BR$(B^0 \rightarrow K^0 (X \rightarrow l^+ l^-))  = (3.51^{+0.69}_{-0.60} \pm 0.10) \times 10^{-7}$ for the entire $q^2$ range \cite{BELLE:2019xld}. They place cuts on the dielectron final state for $B \rightarrow J/\psi K$ at $8.12 <q^2 <10.2$~GeV and for $B\rightarrow \psi(2S) K $ at $12.8 < q^2 < 14.0$~GeV$^2$ an additional veto for the low $q^2$ regime is given by $ q^2 > 0.05~\text{ GeV}^2$ to suppress any contamination from $\gamma^* \rightarrow e^+e^-$ and $\pi^0 \rightarrow \gamma e^+e^-$.

At the LHCb detector, there are ongoing efforts to search for decays of an X boson. Past searches have been focused on the prompt decay of $B^{\pm} \rightarrow K^{\pm} (X \rightarrow \mu^+ \mu^-)$ and have derived a limit on the branching ratio of ($4.36 \pm 0.15 \pm 0.18) \times 10^{-7}$ with $0.05 < q^2 < 22.00~\text{ GeV}^2$ \cite{LHCb:2012juf}. 

The experimental searches at BaBar \cite{BaBar:2008jdv} are summarized in Table \ref{table:BaBaR_values}, and provide further bounds on leptonic channels. These branching ratios are categorized based on $q^2$, stemming from Belle's strategy to exclude meson resonances. Notably, the extended low region, defined by $q^2 < 0.1$ GeV$^2$, is attributed to an amplified coupling to the photonic penguin amplitude exclusive to this specific mode.
BaBar reports branching ratios of $BR(B \rightarrow K l^+ l^-) = (4.7 \pm 0.6 \pm 0.2) \times 10^{-7}$ and $BR(B \rightarrow K^* l^+ l^-) = (10.2^{+1.4}_{-1.3} \pm 0.5) \times 10^{-7}$ for $m^2_{ll} > 0.1$ GeV \cite{BaBar:2012mrf}.

\begin{table}[t]
\centering
\caption{Branching ratios for $B$ meson decays in various intervals of $q^2$, categorized as: extended low $q^2$ ($q^2 < 0.1~\text{GeV}^2$), low $q^2$ ($0.1 < q^2 < 7.02~\text{GeV}^2$), and high $q^2$ ($q^2 > 10.24~\text{GeV}^2$) \cite{BaBar:2008jdv}.}
\begin{tabular}{|c|c|c|c|}
\hline
 & Extended Low q$^2$ & Low q$^2$ & High q$^2$ \\
\hline
$BR(B^+ \rightarrow K^{* +}e^+ e^-)$ & $1.32^{+0.41}_{-0.36} \pm 0.09 \times 10^{-6}$ & $1.06^{+0.31}_{-0.28} \pm 0.07 \times 10^{-6}$ & $0.19^{+0.23}_{-0.21} \pm 0.01 \times 10^{-6}$ \\
\hline
$BR(B^0 \rightarrow K^{* 0}e^+ e^-)$ & $0.73^{+0.22}_{-0.19} \pm 0.04 \times 10^{-6}$ & $0.20^{+0.12}_{-0.11} \pm 0.01 \times 10^{-6}$ & $0.35^{+0.15}_{-0.13} \pm 0.02 \times 10^{-6}$ \\
\hline
\end{tabular}
\label{table:BaBaR_values}
\end{table}

\begin{figure}[t]
    \centering
    \includegraphics[width=\columnwidth]{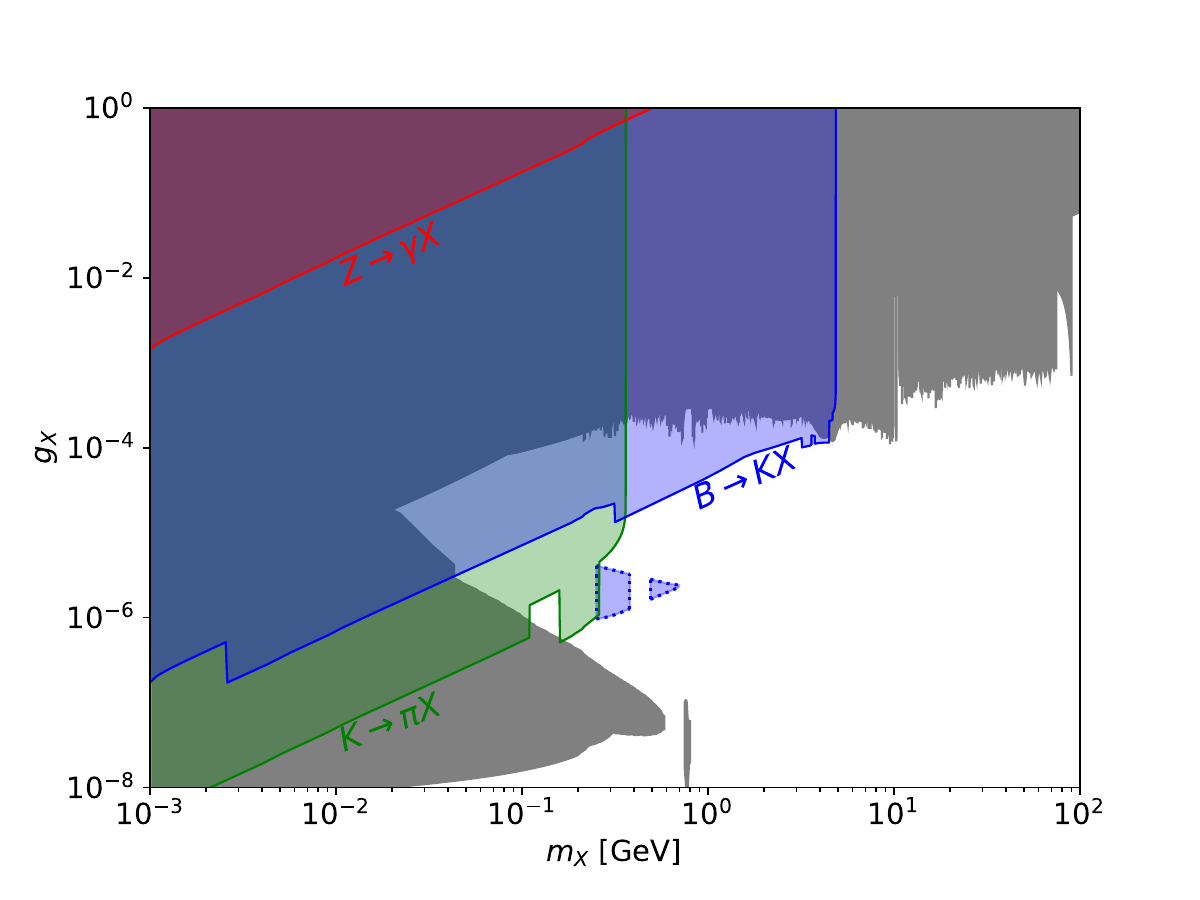}
    \caption{Constraints on the parameter space of $U(1)_{B-L}$, in the plane of the boson mass $m_X$ and its coupling $g_X$.
    Shaded gray areas represent existing experimental constraints, encompassing resonance searches, beam dumps, and neutrino scattering experiments~\cite{A1:2011yso,Merkel:2014avp,Bodas:2021fsy,Abrahamyan:2011gv,Ablikim:2017aab,Lees:2014xha,TheBABAR:2016rlg,Bauer:2018onh,Bergsma:1985qz,Tsai:2019buq, CMS:2019foo, Andreas:2012mt,Bjorken:2009mm,Riordan:1987aw,Bross:1989mp,Petersen:2023hgm,FASER:2023tle,Adrian:2018scb,Konaka:1986cb,Anastasi:2015qla,Anastasi:2016ktq,Anastasi:2018azp,Babusci:2012cr,Babusci:2014sta,Aaij:2017rft,Aaij:2019bvg,Batley:2015lha,Dobrich:2023dkm,Banerjee:2018vgk,Banerjee:2019hmi,Astier:2001ck,Blumlein:1990ay,Blumlein:1991xh,Davier:1989wz,Bernardi:1985ny} via DarkCast \cite{Ilten:2018crw}. The shaded red area corresponds to new constraints we derive from LEP searches for $Z \rightarrow \gamma + X(\rightarrow \text{inv})$, blue and green shading indicates constraints derived from from $B$ and $K$ meson decays.    
    \label{fig:reachplot}}
\end{figure}

In addition, there are displaced vertex searches at LHCb where they specifically focus on $B^+ \rightarrow K^{+} (X\rightarrow \mu^+ \mu^-)$ \cite{LHCb:2015nkv}. Upper limits are placed in the mass range $250 < m_X / \text{MeV} < 4700$ for lifetimes within the range $0.1 < \tau_X / \text{ps} < 1000$. The branching ratio falls between $2 \times 10^{-10} $ and $10^{-7}$, illustrated in their Figure 4. We adopt a conservative approach, ensuring the mass and couplings under consideration fall within $1 < \tau_X / \text{ps} < 100$ with a branching ratio $< 10^{-9}$. This leads to the wedge shaped constraint in Fig.~\ref{fig:reachplot}. 

Lastly, Kaon decays can provide some of the leading constraints over a wide swath of parameter space. A recent effort by the NA62 experiment set the upper limits on the branching ratio at $< 3-6 \times 10^{-11}$ in the mass range 0-110 MeV and $< 1 \times 10^{-11}$ between 160-260 MeV for the decay process $K^+ \rightarrow \pi^+ (X\rightarrow \text{inv})$  \cite{NA62:2021zjw}. We conservatively adopt the lower limit $ 3 \times 10^{-11}$ as our constraint.  An additional search from E949, which combines its data with the preceding experiment E787, established a limit on BR$(K^+ \rightarrow \pi^+ \nu \Bar{\nu}) = 1.73^{+ 1.15}_{- 1.05} \times 10^{-10}$ \cite{E949:2008btt}, which is a signature for an invisibly decaying $B-L$ boson. 

Visible decays are constrained by recent results from NA62 improving upon prior NA48/2 constraints. Specifically, they constrain BR$(K^\pm \rightarrow \pi^\pm \mu^+\mu^-) = 9.15 \pm 0.08 \times 10^{-8}$ \cite{NA62:2022qes}. Furthermore, there are stronger constraints from the KTeV/E799 experiment which looks at $K_L \rightarrow \pi^0 (X \rightarrow e^+ e^-) $ for $m_{ee} > m_{\pi^0}$, a restriction imposed due to backgrounds from Dalitz decays of $\pi^0$ in processes such as $K_L \rightarrow \pi^0 \pi^0$ and $K_L \rightarrow \pi^0 \pi^0 \pi^0$. \cite{KTeV:2003sls}. Complementary work from KTeV establishes an upper bound on the branching ratio BR$(K_L \rightarrow \pi^0 (X \rightarrow \mu^+ \mu^-)) < 3.8 \times 10^{-10}$ \cite{KTEV:2000ngj}.

The impact of all of these searches is summarized in Figure~\ref{fig:reachplot} in the plane of the boson mass $m_X$ and its coupling $g_X$. We assume negligible kinetic mixing with SM hypercharge.  Grey shading on the figure indicates pre-existing constraints by DarkCast \cite{Ilten:2018crw} from a wide variety of measurements, including resonance searches, beam dumps, and neutrino scattering experiments~\cite{A1:2011yso,Merkel:2014avp,Bodas:2021fsy,Abrahamyan:2011gv,Ablikim:2017aab,Lees:2014xha,TheBABAR:2016rlg,Bauer:2018onh,Bergsma:1985qz,Tsai:2019buq, CMS:2019foo, Andreas:2012mt,Bjorken:2009mm,Riordan:1987aw,Bross:1989mp,Petersen:2023hgm,FASER:2023tle,Adrian:2018scb,Konaka:1986cb,Anastasi:2015qla,Anastasi:2016ktq,Anastasi:2018azp,Babusci:2012cr,Babusci:2014sta,Aaij:2017rft,Aaij:2019bvg,Batley:2015lha,Dobrich:2023dkm,Banerjee:2018vgk,Banerjee:2019hmi,Astier:2001ck,Blumlein:1990ay,Blumlein:1991xh,Davier:1989wz,Bernardi:1985ny}.  Constraints derived in this work based on combined searches based on $Z$, $B$, and $K$ decays are shown as regions shaded in red, blue, and green, respectively.  Evident from the figure, while $Z$ decays constrain no additional parameter space, $B$ and $K$ decays provide significant improvements, spanning several orders of magnitude in the mass range from $\sim$ 20 MeV to $\sim$ 2 GeV. The small island of constraints from $B$ meson decays corresponds to the displaced vertex search carried out by LHCb\cite{LHCb:2015nkv}.

\section{Constraints on an Intergenerational B-L}
\label{sec:b3-l2}

While our analysis has focused on a gauge boson coupled to $U(1)_{B-L}$, it applies to any $U(1)$ extension under which the Standard Model fermions are vector-like.
Intergenerational $B_i - L_j$ gauge symmetries are also anomaly free within the Standard Model, and appear widely in popular extensions. We specifically examine the model of $U(1)_{B_3-L_2}$ considered in Ref.~\cite{Okada:2023mdv} which extends the SM by a gauged $U(1)_{B_3-L_2}$ accompanied by three RH neutrinos, at least one of which carries charge $B_3 - L_2 = -1$ to cancel the $U(1)^3_{B_3-L_2}$ anomaly.  The remaining two have arbitrary equal and opposite charges $x_N$ and $-x_N$, respectively. The two RH neutrinos carrying charges $\pm x_N$ combine into a Dirac spinor which can play the role of dark matter:
\begin{equation}
    \chi = 
    \begin{pmatrix}
    \nu_R^3\\
    \nu_R^{i *}
    \end{pmatrix} ~.
\end{equation}

\begin{figure}
    \centering
    \includegraphics[width=\columnwidth]{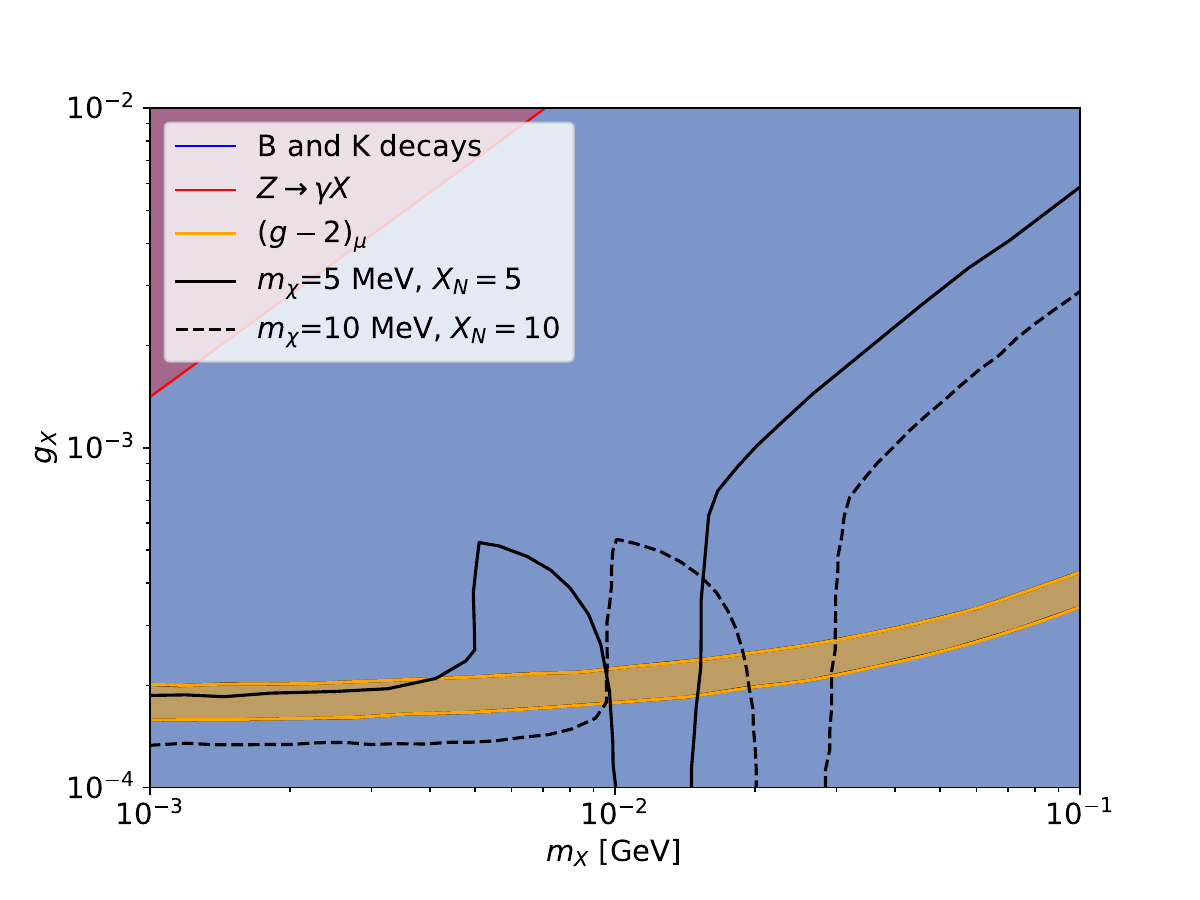}
    \caption{The parameter space of the mass and coupling of a $U(1)_{B_3 - L_2}$ gauge boson, as discussed in Ref.~\cite{Okada:2023mdv}.  The solid (dashed) black lines correspond to the correct DM relic density DM mass of $m_\chi = 5 (10) MeV$ and $B_3 -L_2$ charge $x_N = 5 (10)$. The orange band represents the region favored to explain the $(g-2)_\mu$ observations. The mass of the RH neutrino is chosen to be $m_{\nu_R} = 0.1$ GeV. The blue region (which covers the entire plot) indicates the constraints from $B$ and $K$ meson decays, ruling out the favored region of parameter space.}
    \label{fig:b3-l2}
\end{figure}

The authors of \cite{Okada:2023mdv} select specific masses for the RH neutrino and $\chi$ to account for the correct relic density of DM via freeze-out, while simultaneously explaining the observed discrepancy between the measured $(g-2)_\mu$ and its SM prediction. Assuming negligible kinetic mixing compared to the gauge coupling $g_X$, we demonstrate in Figure~\ref{fig:b3-l2} that the region of parameter space that correctly describes the relic density and $(g-2)_\mu$ (for two choices of DM mass and charge) is robustly ruled out by the kaon and $B$ meson decays discussed above. While these results correspond to specific choices of DM and RH neutrino masses and couplings, they illustrate the important point that there can be important constraints on non-anomalous $U(1)$ theories from processes mediated by triangle amplitudes.

\section{Discussion}
\label{sec:discussion}

In this work, we show that while conserved currents are safe from anomalies in the UV, the chiral electroweak interactions combined with electroweak symmetry-breaking can result in non-zero triangle amplitudes in the IR. In particular, we examine $U(1)_{B-L}$ and find that processes operating at energies below the mass of the top quark, including $B$ and $K$ meson decays provide key constraints on  regions of parameter space that would otherwise be allowed. Our results for $B-L$ are summarized in Figure \ref{fig:reachplot}. We also study the case of an intergenerational $B_3 -L_2$ MeV model of DM \cite{Okada:2023mdv} which seeks to solve the $(g-2)_\mu$ anomaly, and find that the constraints arising from the triangle vertices completely rule out the region of parameter space which would resolve both DM and $(g-2)_\mu$. 

Our results are far more general than these specific examples, and apply to any $U(1)$ gauge extension of the SM under which the SM fermions are charged.

\section{Acknowledgements}
TMPT is grateful for conversations with Susan Gardner.  TBS is supported by the National Science Foundation Graduate Research Fellowship Program under Grant No. 1839285.  The work of TMPT is supported in part by the U.S.\ National Science Foundation under Grant PHY-2210283. 

\appendix 

\section{Triangle Amplitudes}
\label{app:anomalyamps}

In this appendix, we derive the divergence of the triangle amplitudes in Fig. \ref{fig:XBBfeynmandiagram}.  The corresponding amplitude for $X$ coupled to two hypercharge bosons is:
\begin{equation}
 \begin{split}
    \mathcal{M}^{\mu \nu \rho} = (-1) \sum_{f} \int \frac{d^4 l}{(2 \pi)^4} 
    \Bigg\{ Tr\left[ (i g_X X \gamma^\mu ) 
    \frac{i(\slashed{l} - \slashed{q} +m)}{(l-q)^2 -m^2} 
    g_B^\rho \frac{i (\slashed{l} + m)}{l^2 -m^2} g_B^\nu
    \frac{i(\slashed{l} + \slashed{p} +m)}{(l+p)^2 -m^2}
    \right]
    \\
     + Tr\left[ (i g_X X \gamma^\mu ) 
    \frac{i(\slashed{l} - \slashed{p} +m)}{(l-p)^2 -m^2} 
    g_B^\nu \frac{i (\slashed{l} + m)}{l^2 -m^2} g_B^\rho
    \frac{i(\slashed{l} + \slashed{q} +m)}{(l+q)^2 -m^2}
    \right]\Bigg\} ~,
 \end{split}
\end{equation}
where $g_B^{\rho,\nu} \equiv i( g_V \gamma^{\rho,\nu} + g_A \gamma^{\rho,\nu} \gamma^5)$ represents the coupling of the hypercharge boson to $f$, described in terms of its vector and axial vector components: $g_A = g_1 \frac{(Y_R + Y_L)}{2}$ and $g_V = g_1 \frac{(Y_R - Y_L)}{2}$ with $Y_{R,L}$ denoting the hypercharge of the fermion circulating in the loop. 
Our focus is on the longitudinal amplitude $(p+q)_\mu \mathcal{M}^{\mu \nu \rho} $, thus we replace the appropriate vertices with $(\slashed{p}+\slashed{q})$ and apply the identity: $(\slashed{p}+\slashed{q}) = (\slashed{l}+\slashed{p}-m) - (\slashed{l}-\slashed{q} -m) = (\slashed{l}+\slashed{q}-m) - (\slashed{l}-\slashed{p}-m)$ leading to:
\begin{equation}
 \begin{split}
    &(p+q)_\mu\mathcal{M}^{\mu \nu \rho} = \sum_f g_X g_V g_A Tr[X] \times \\
    &-Tr\left[ \gamma^\rho \frac{(\slashed{l} +m)}{l^2 -m^2} \gamma^\nu \gamma_5 
    \frac{(\slashed{l} + \slashed{p} + m)}{(l+p)^2 -m^2} 
    \right]
    + Tr\left[ \frac{(\slashed{l} - \slashed{q} + m)}{(l+q)^2 -m^2} \gamma^\rho \frac{(\slashed{l} +m)}{l^2 -m^2} \gamma^\nu \gamma_5 
    \right] \\
    &-Tr\left[ \gamma^\rho \gamma_5 \frac{(\slashed{l} +m)}{l^2 -m^2} \gamma^\nu  
    \frac{(\slashed{l} + \slashed{p} + m)}{(l+p)^2 -m^2} 
    \right] 
    + Tr\left[ \frac{(\slashed{l} - \slashed{q} + m)}{(l+q)^2 -m^2} \gamma^\rho \gamma_5 \frac{(\slashed{l} +m)}{l^2 -m^2} \gamma^\nu 
    \right]  +  \hspace{0.2cm} 
    \Bigg[\substack{\rho \leftrightarrow \nu \vspace{0.2cm}  \\ p \leftrightarrow q }\Bigg] ~.
 \end{split}
\end{equation}

It is convenient to apply a shift of the form $l \rightarrow l + a$ with $a$ chosen to cause the terms cancel in pairs. However, this operation is subtle, since the integrals are linearly divergent and such a momentum shift results in a finite change in the integral $\Delta ^{\mu \nu \rho } (a)$, proportional to the shift \cite{Raby2022,Schwartz:2014sze}. We adopt an arbitrary momentum shift $a$, which we determine by requiring that the final result satisfy the Ward identities. The finite result is then given by:
\begin{equation}
    \Delta ^{\mu \nu \rho } (a) = \mathcal{M} ^{\mu \nu \rho } (a) - \mathcal{M} ^{\mu \nu \rho } (0) ~.
\end{equation}

Thus, we focus on the uncrossed diagram contribution $\Delta ^{\mu \nu \rho }_1$, where there is a comparable finite contribution from the crossed diagram $\Delta ^{\mu \nu \rho }_2$. Taylor expanding around a=0 and keeping only the leading order terms (since the higher order derivatives vanish too quickly at infinity to contribute), we obtain:
\begin{equation}
\begin{split}
    \Delta ^{\mu \nu \rho }_1 &= \sum_f g_X g_V g_A X \int \frac{d^4 l}{(2 \pi)^4} a^\tau \frac{\partial}{\partial l^\tau} Tr\Bigg[ \gamma^\mu \frac{(\slashed{l} - \slashed{q} +m)}{(l-q)^2 -m^2} \gamma^\rho
    \frac{(\slashed{l} + m)}{l^2 -m^2} \gamma^\nu \gamma_5 \frac{(\slashed{l} + \slashed{p} +m)}{(l+p)^2 -m^2}
    \Bigg] \\
    &= \left(\sum_f g_X g_V g_A X \right) \frac{a^\tau}{(2\pi)^4} \lim\limits_{l \to \infty} \int d\Omega l_\tau l^2 \frac{Tr[\gamma^\mu \slashed{l}\gamma^\rho \slashed{l}\gamma^\nu \gamma^5 \slashed{l}]}{l^6} \\
    &= \left(\sum_f g_X g_V g_A X \right)\frac{-2 \pi^2 a^\tau}{(2 \pi)^4}\lim\limits_{l \to \infty}  4 i \epsilon^{\alpha \rho \nu \mu} \frac{g_{\tau \alpha}}{4} ~.
\end{split}
\end{equation}

Summing contributions from the swapped diagram and explicitly selecting the shift as an arbitrary linear combination of the momenta $a = u q + v p$:
\begin{equation}
\begin{split}
    \Delta ^{\mu \nu \rho } &= \left(\sum_f g_X g_V g_A X \right) \frac{i}{4 \pi^2} [(u q_\alpha + v p_\alpha) + (u p_\alpha + v q_\alpha )]\epsilon^{\alpha \mu \nu \rho} ~. \\
\end{split}
\end{equation}
We define $\Hat{u}=u-v$ and write the longitudinal contribution as:
\begin{equation}
    (p+q)_\mu\mathcal{M}^{\mu \nu \rho} (\Hat{u}) = \sum_f   \frac{i g_X g_V g_A X}{4 \pi^2} \Hat{u} (p+q)_\mu (q-p)_\alpha \epsilon^{\alpha \mu \nu \rho}
\end{equation}

We determine the shift $\Hat{u}$ by imposing the vector ward identities for $B_\nu$ and $ B_\rho$. To uphold the gauge invariance associated with the hypercharge bosons, the shift must take the form $\Hat{u} = 2 - 4 m_f^2 I_{00}$ where $I_{00}$ is given by:
\begin{equation}
    I_{00}(p, q, m_f) \equiv \int^1_0 dx \int^{1-x}_0 \frac{dz}{m_f^2 - z(1-z)p^2 - x(1-x)q^2 - 2xz (p \cdot q)} .
\end{equation}

All together, the triangle vertices for $U(1)_Y^2 U(1)_X$ and $SU(2)_L^2 U(1)_X$ are given by:
\begin{eqnarray}
    (p+q)_\mu \mathcal{M}^{\mu \nu \rho}_{XBB} & = & \sum_f \frac{i g_X g_V g_A }{\pi^2} X 
    ~ \left[ 1 - 2m_f^2 I_{00}(p, q, m_f) \right] 
    ~ \epsilon^{\alpha \beta \nu \rho} q_\alpha p_\beta , \\
    (p+q)_\mu \mathcal{M}^{\mu \nu \rho}_{XWW} & = & -\frac{1}{4}\sum_f \frac{i g_X g_2^2}{\pi^2}  \Tr [(X)T^aT^b]
    ~ \left[ 1 - 2m_f^2 I_{00}(p, q, m_f) \right]
    ~ \epsilon^{\alpha \beta \nu \rho} q_\alpha p_\beta .
\end{eqnarray}

\section{Form Factors}
\label{app:formfacs}

In this appendix, we highlight the form factors used in equations \ref{eq:BtoKX},\ref{eq:BtoKstarX},\ref{eq:KtopiX},\ref{eq:KLtopi0X} from \cite{Ball:2004rg,Ball:2004ye}. For $B \rightarrow K^*$, the hadronic matrix element is \cite{Ball:2004rg}:
\begin{equation}
\begin{split}
    \bra{K^*(p)} \Bar{q} \gamma_\mu (1 - \gamma_5) b \ket{B(p_B)} = &-i \epsilon^*_\mu (m_B+m_{K^*}) A_1(q^2) + i(p_B + p)_\mu (\epsilon^* q) \frac{A_2(q^2)}{(m_B + m_{K^*})} \\
    &+ i q_\mu (\epsilon^* q) \frac{2m_V}{q^2} [A_3 (q^2) - A_0(q^2)] 
    + \epsilon^{\mu \nu \rho \sigma} \epsilon^{* \nu} p^\rho_B p^\sigma \frac{2 V(q^2)}{m_B +m_{K^*}} ,
\end{split}
\end{equation}
where in this case $V(p)$ is $K^*$ and $q$ is the momentum of the $X$ boson. Contracting this with the outgoing $X$ boson we are left with a dependence only on $A_3(q^2)$ where at zero momentum transfer we have $A_3(0)=A_0(0)$, but more generally:
\begin{equation}
    A_3(q^2) = \frac{m_B + m_{K^*}}{2 m_{K^*}} A_1(q^2) - \frac{m_B - m_{K^*}}{2 m_{K^*}} A_2(q^2)
\end{equation}
\begin{equation}
    A_1(q^2) = \frac{0.290}{1-q^2/(40.38 \text{GeV}^2)}
\end{equation}
\begin{equation}
    A_2(q^2) = \frac{-0.084}{1-q^2 / (52.0 \text{ GeV}^2)} +  \frac{0.342}{(1-q^2 / (52.0 \text{ GeV}^2))^2} .
\end{equation}


Similarly, for the case of $B \rightarrow K$, the hadronic matrix element is parametrized\cite{Ball:2004ye}:
\begin{equation}
    \bra{K(p)} \Bar{u} \gamma_\mu b \ket{B(p_B)} = \left\{ (p + p_B)_\mu - \frac{m_B^2 - m_K^2}{q^2} q_\mu \right\} f_+^K(q^2) + \left\{\frac{m_B^2 - m_K^2}{q^2} q_\mu 
    \right\}f_0^K(q^2)
\end{equation}
Under our approximations, after contracting with the external $X$ boson's longitudinal polarization, the only dependence is on $f_+^K(q^2)$. We note that although $f_+^K(0) = f_0^K(0)$, this relation does not hold for the generalization to larger $q^2$. Thus, the correct choice is $f_+^K(q^2)$, using the values from set 2 in \cite{Ball:2004ye}:
\begin{equation}
    f_+^K (q^2) = \frac{0.1616}{1-q^2 / (29.3 \text{ GeV}^2)} +  \frac{0.1730}{1-q^2 / (29.3 \text{ GeV}^2)} .
\end{equation}

\bibliography{mybib}

\providecommand{\noopsort}[1]{}\providecommand{\singleletter}[1]{#1}%
\begin{thebibliography}{59}%
\makeatletter
\providecommand \@ifxundefined [1]{%
 \@ifx{#1\undefined}
}%
\providecommand \@ifnum [1]{%
 \ifnum #1\expandafter \@firstoftwo
 \else \expandafter \@secondoftwo
 \fi
}%
\providecommand \@ifx [1]{%
 \ifx #1\expandafter \@firstoftwo
 \else \expandafter \@secondoftwo
 \fi
}%
\providecommand \natexlab [1]{#1}%
\providecommand \enquote  [1]{``#1''}%
\providecommand \bibnamefont  [1]{#1}%
\providecommand \bibfnamefont [1]{#1}%
\providecommand \citenamefont [1]{#1}%
\providecommand \href@noop [0]{\@secondoftwo}%
\providecommand \href [0]{\begingroup \@sanitize@url \@href}%
\providecommand \@href[1]{\@@startlink{#1}\@@href}%
\providecommand \@@href[1]{\endgroup#1\@@endlink}%
\providecommand \@sanitize@url [0]{\catcode `\\12\catcode `\$12\catcode
  `\&12\catcode `\#12\catcode `\^12\catcode `\_12\catcode `\%12\relax}%
\providecommand \@@startlink[1]{}%
\providecommand \@@endlink[0]{}%
\providecommand \url  [0]{\begingroup\@sanitize@url \@url }%
\providecommand \@url [1]{\endgroup\@href {#1}{\urlprefix }}%
\providecommand \urlprefix  [0]{URL }%
\providecommand \Eprint [0]{\href }%
\providecommand \doibase [0]{https://doi.org/}%
\providecommand \selectlanguage [0]{\@gobble}%
\providecommand \bibinfo  [0]{\@secondoftwo}%
\providecommand \bibfield  [0]{\@secondoftwo}%
\providecommand \translation [1]{[#1]}%
\providecommand \BibitemOpen [0]{}%
\providecommand \bibitemStop [0]{}%
\providecommand \bibitemNoStop [0]{.\EOS\space}%
\providecommand \EOS [0]{\spacefactor3000\relax}%
\providecommand \BibitemShut  [1]{\csname bibitem#1\endcsname}%
\let\auto@bib@innerbib\@empty
\bibitem [{\citenamefont {Dror}\ \emph
  {et~al.}(2017{\natexlab{a}})\citenamefont {Dror}, \citenamefont {Lasenby},\
  and\ \citenamefont {Pospelov}}]{Dror2017a}%
  \BibitemOpen
  \bibfield  {author} {\bibinfo {author} {\bibfnamefont {J.~A.}\ \bibnamefont
  {Dror}}, \bibinfo {author} {\bibfnamefont {R.}~\bibnamefont {Lasenby}},\ and\
  \bibinfo {author} {\bibfnamefont {M.}~\bibnamefont {Pospelov}},\ }\bibfield
  {title} {\bibinfo {title} {{New constraints on light vectors coupled to
  anomalous currents}},\ }\href
  {https://doi.org/10.1103/PhysRevLett.119.141803} {\bibfield  {journal}
  {\bibinfo  {journal} {Phys. Rev. Lett.}\ }\textbf {\bibinfo {volume} {119}},\
  \bibinfo {pages} {141803} (\bibinfo {year} {2017}{\natexlab{a}})},\ \Eprint
  {https://arxiv.org/abs/1705.06726} {arXiv:1705.06726 [hep-ph]} \BibitemShut
  {NoStop}%
\bibitem [{\citenamefont {Dror}\ \emph
  {et~al.}(2017{\natexlab{b}})\citenamefont {Dror}, \citenamefont {Lasenby},\
  and\ \citenamefont {Pospelov}}]{Dror2017b}%
  \BibitemOpen
  \bibfield  {author} {\bibinfo {author} {\bibfnamefont {J.~A.}\ \bibnamefont
  {Dror}}, \bibinfo {author} {\bibfnamefont {R.}~\bibnamefont {Lasenby}},\ and\
  \bibinfo {author} {\bibfnamefont {M.}~\bibnamefont {Pospelov}},\ }\bibfield
  {title} {\bibinfo {title} {{Dark forces coupled to nonconserved currents}},\
  }\href {https://doi.org/10.1103/PhysRevD.96.075036} {\bibfield  {journal}
  {\bibinfo  {journal} {Phys. Rev. D}\ }\textbf {\bibinfo {volume} {96}},\
  \bibinfo {pages} {075036} (\bibinfo {year} {2017}{\natexlab{b}})},\ \Eprint
  {https://arxiv.org/abs/1707.01503} {arXiv:1707.01503 [hep-ph]} \BibitemShut
  {NoStop}%
\bibitem [{\citenamefont {Cornwall}\ \emph {et~al.}(1974)\citenamefont
  {Cornwall}, \citenamefont {Levin},\ and\ \citenamefont
  {Tiktopoulos}}]{Cornwall:1974km}%
  \BibitemOpen
  \bibfield  {author} {\bibinfo {author} {\bibfnamefont {J.~M.}\ \bibnamefont
  {Cornwall}}, \bibinfo {author} {\bibfnamefont {D.~N.}\ \bibnamefont
  {Levin}},\ and\ \bibinfo {author} {\bibfnamefont {G.}~\bibnamefont
  {Tiktopoulos}},\ }\bibfield  {title} {\bibinfo {title} {{Derivation of Gauge
  Invariance from High-Energy Unitarity Bounds on the s Matrix}},\ }\href
  {https://doi.org/10.1103/PhysRevD.10.1145} {\bibfield  {journal} {\bibinfo
  {journal} {Phys. Rev. D}\ }\textbf {\bibinfo {volume} {10}},\ \bibinfo
  {pages} {1145} (\bibinfo {year} {1974})},\ \bibinfo {note} {[Erratum:
  Phys.Rev.D 11, 972 (1975)]}\BibitemShut {NoStop}%
\bibitem [{\citenamefont {Ball}\ and\ \citenamefont
  {Zwicky}(2005{\natexlab{a}})}]{Ball:2004rg}%
  \BibitemOpen
  \bibfield  {author} {\bibinfo {author} {\bibfnamefont {P.}~\bibnamefont
  {Ball}}\ and\ \bibinfo {author} {\bibfnamefont {R.}~\bibnamefont {Zwicky}},\
  }\bibfield  {title} {\bibinfo {title} {{$B_{d,s} \to \rho, \omega, K^*, \phi$
  decay form-factors from light-cone sum rules revisited}},\ }\href
  {https://doi.org/10.1103/PhysRevD.71.014029} {\bibfield  {journal} {\bibinfo
  {journal} {Phys. Rev. D}\ }\textbf {\bibinfo {volume} {71}},\ \bibinfo
  {pages} {014029} (\bibinfo {year} {2005}{\natexlab{a}})},\ \Eprint
  {https://arxiv.org/abs/hep-ph/0412079} {arXiv:hep-ph/0412079} \BibitemShut
  {NoStop}%
\bibitem [{\citenamefont {Ball}\ and\ \citenamefont
  {Zwicky}(2005{\natexlab{b}})}]{Ball:2004ye}%
  \BibitemOpen
  \bibfield  {author} {\bibinfo {author} {\bibfnamefont {P.}~\bibnamefont
  {Ball}}\ and\ \bibinfo {author} {\bibfnamefont {R.}~\bibnamefont {Zwicky}},\
  }\bibfield  {title} {\bibinfo {title} {{New results on $B \to \pi, K, \eta$
  decay formfactors from light-cone sum rules}},\ }\href
  {https://doi.org/10.1103/PhysRevD.71.014015} {\bibfield  {journal} {\bibinfo
  {journal} {Phys. Rev. D}\ }\textbf {\bibinfo {volume} {71}},\ \bibinfo
  {pages} {014015} (\bibinfo {year} {2005}{\natexlab{b}})},\ \Eprint
  {https://arxiv.org/abs/hep-ph/0406232} {arXiv:hep-ph/0406232} \BibitemShut
  {NoStop}%
\bibitem [{\citenamefont {Acciarri}\ \emph {et~al.}(1997)\citenamefont
  {Acciarri} \emph {et~al.}}]{L3:1997exg}%
  \BibitemOpen
  \bibfield  {author} {\bibinfo {author} {\bibfnamefont {M.}~\bibnamefont
  {Acciarri}} \emph {et~al.} (\bibinfo {collaboration} {L3}),\ }\bibfield
  {title} {\bibinfo {title} {{Search for new physics in energetic single photon
  production in $e^{+} e^{-}$ annihilation at the $Z$ resonance}},\ }\href
  {https://doi.org/10.1016/S0370-2693(97)01003-4} {\bibfield  {journal}
  {\bibinfo  {journal} {Phys. Lett. B}\ }\textbf {\bibinfo {volume} {412}},\
  \bibinfo {pages} {201} (\bibinfo {year} {1997})}\BibitemShut {NoStop}%
\bibitem [{\citenamefont {Rind}(1992)}]{Rind:1992es}%
  \BibitemOpen
  \bibfield  {author} {\bibinfo {author} {\bibfnamefont {O.}~\bibnamefont
  {Rind}} (\bibinfo {collaboration} {L3}),\ }\bibfield  {title} {\bibinfo
  {title} {{Isolated hard photon emission in hadronic Z decays}},\ }in\
  \href@noop {} {\emph {\bibinfo {booktitle} {{7th Meeting of the APS Division
  of Particles Fields}}}}\ (\bibinfo {year} {1992})\ pp.\ \bibinfo {pages}
  {961--964}\BibitemShut {NoStop}%
\bibitem [{\citenamefont {Acton}\ \emph {et~al.}(1991)\citenamefont {Acton}
  \emph {et~al.}}]{OPAL:1991acn}%
  \BibitemOpen
  \bibfield  {author} {\bibinfo {author} {\bibfnamefont {P.~D.}\ \bibnamefont
  {Acton}} \emph {et~al.} (\bibinfo {collaboration} {OPAL}),\ }\bibfield
  {title} {\bibinfo {title} {{A Measurement of photon radiation in lepton pair
  events from Z0 decays}},\ }\href
  {https://doi.org/10.1016/0370-2693(91)91694-Q} {\bibfield  {journal}
  {\bibinfo  {journal} {Phys. Lett. B}\ }\textbf {\bibinfo {volume} {273}},\
  \bibinfo {pages} {338} (\bibinfo {year} {1991})}\BibitemShut {NoStop}%
\bibitem [{\citenamefont {Adeva}\ \emph {et~al.}(1991)\citenamefont {Adeva}
  \emph {et~al.}}]{L3:1991kow}%
  \BibitemOpen
  \bibfield  {author} {\bibinfo {author} {\bibfnamefont {B.}~\bibnamefont
  {Adeva}} \emph {et~al.} (\bibinfo {collaboration} {L3}),\ }\bibfield  {title}
  {\bibinfo {title} {{Search for narrow high mass resonances in radiative
  decays of the Z0}},\ }\href {https://doi.org/10.1016/0370-2693(91)90659-E}
  {\bibfield  {journal} {\bibinfo  {journal} {Phys. Lett. B}\ }\textbf
  {\bibinfo {volume} {262}},\ \bibinfo {pages} {155} (\bibinfo {year}
  {1991})}\BibitemShut {NoStop}%
\bibitem [{\citenamefont {Grygier}\ \emph {et~al.}(2017)\citenamefont {Grygier}
  \emph {et~al.}}]{Belle:2017oht}%
  \BibitemOpen
  \bibfield  {author} {\bibinfo {author} {\bibfnamefont {J.}~\bibnamefont
  {Grygier}} \emph {et~al.} (\bibinfo {collaboration} {Belle}),\ }\bibfield
  {title} {\bibinfo {title} {{Search for $\boldsymbol{B\to h\nu\bar{\nu}}$
  decays with semileptonic tagging at Belle}},\ }\href
  {https://doi.org/10.1103/PhysRevD.96.091101} {\bibfield  {journal} {\bibinfo
  {journal} {Phys. Rev. D}\ }\textbf {\bibinfo {volume} {96}},\ \bibinfo
  {pages} {091101} (\bibinfo {year} {2017})},\ \bibinfo {note} {[Addendum:
  Phys.Rev.D 97, 099902 (2018)]},\ \Eprint {https://arxiv.org/abs/1702.03224}
  {arXiv:1702.03224 [hep-ex]} \BibitemShut {NoStop}%
\bibitem [{\citenamefont {Choudhury}\ \emph {et~al.}(2021)\citenamefont
  {Choudhury} \emph {et~al.}}]{BELLE:2019xld}%
  \BibitemOpen
  \bibfield  {author} {\bibinfo {author} {\bibfnamefont {S.}~\bibnamefont
  {Choudhury}} \emph {et~al.} (\bibinfo {collaboration} {BELLE}),\ }\bibfield
  {title} {\bibinfo {title} {{Test of lepton flavor universality and search for
  lepton flavor violation in $B \rightarrow K\ell \ell$ decays}},\ }\href
  {https://doi.org/10.1007/JHEP03(2021)105} {\bibfield  {journal} {\bibinfo
  {journal} {JHEP}\ }\textbf {\bibinfo {volume} {03}},\ \bibinfo {pages}
  {105}},\ \Eprint {https://arxiv.org/abs/1908.01848} {arXiv:1908.01848
  [hep-ex]} \BibitemShut {NoStop}%
\bibitem [{\citenamefont {Aaij}\ \emph {et~al.}(2013)\citenamefont {Aaij} \emph
  {et~al.}}]{LHCb:2012juf}%
  \BibitemOpen
  \bibfield  {author} {\bibinfo {author} {\bibfnamefont {R.}~\bibnamefont
  {Aaij}} \emph {et~al.} (\bibinfo {collaboration} {LHCb}),\ }\bibfield
  {title} {\bibinfo {title} {{Differential branching fraction and angular
  analysis of the $B^{+} \rightarrow K^{+}\mu^{+}\mu^{-}$ decay}},\ }\href
  {https://doi.org/10.1007/JHEP02(2013)105} {\bibfield  {journal} {\bibinfo
  {journal} {JHEP}\ }\textbf {\bibinfo {volume} {02}},\ \bibinfo {pages}
  {105}},\ \Eprint {https://arxiv.org/abs/1209.4284} {arXiv:1209.4284 [hep-ex]}
  \BibitemShut {NoStop}%
\bibitem [{\citenamefont {Aubert}\ \emph {et~al.}(2009)\citenamefont {Aubert}
  \emph {et~al.}}]{BaBar:2008jdv}%
  \BibitemOpen
  \bibfield  {author} {\bibinfo {author} {\bibfnamefont {B.}~\bibnamefont
  {Aubert}} \emph {et~al.} (\bibinfo {collaboration} {BaBar}),\ }\bibfield
  {title} {\bibinfo {title} {{Direct CP, Lepton Flavor and Isospin Asymmetries
  in the Decays $B \to K^{(*)} \ell^{+} \ell^{-}$}},\ }\href
  {https://doi.org/10.1103/PhysRevLett.102.091803} {\bibfield  {journal}
  {\bibinfo  {journal} {Phys. Rev. Lett.}\ }\textbf {\bibinfo {volume} {102}},\
  \bibinfo {pages} {091803} (\bibinfo {year} {2009})},\ \Eprint
  {https://arxiv.org/abs/0807.4119} {arXiv:0807.4119 [hep-ex]} \BibitemShut
  {NoStop}%
\bibitem [{\citenamefont {Lees}\ \emph {et~al.}(2012)\citenamefont {Lees} \emph
  {et~al.}}]{BaBar:2012mrf}%
  \BibitemOpen
  \bibfield  {author} {\bibinfo {author} {\bibfnamefont {J.~P.}\ \bibnamefont
  {Lees}} \emph {et~al.} (\bibinfo {collaboration} {BaBar}),\ }\bibfield
  {title} {\bibinfo {title} {{Measurement of Branching Fractions and Rate
  Asymmetries in the Rare Decays $B \to K^{(*)} l^+ l^-$}},\ }\href
  {https://doi.org/10.1103/PhysRevD.86.032012} {\bibfield  {journal} {\bibinfo
  {journal} {Phys. Rev. D}\ }\textbf {\bibinfo {volume} {86}},\ \bibinfo
  {pages} {032012} (\bibinfo {year} {2012})},\ \Eprint
  {https://arxiv.org/abs/1204.3933} {arXiv:1204.3933 [hep-ex]} \BibitemShut
  {NoStop}%
\bibitem [{\citenamefont {Aaij}\ \emph {et~al.}(2015)\citenamefont {Aaij} \emph
  {et~al.}}]{LHCb:2015nkv}%
  \BibitemOpen
  \bibfield  {author} {\bibinfo {author} {\bibfnamefont {R.}~\bibnamefont
  {Aaij}} \emph {et~al.} (\bibinfo {collaboration} {LHCb}),\ }\bibfield
  {title} {\bibinfo {title} {{Search for hidden-sector bosons in $B^0 \!\to
  K^{*0}\mu^+\mu^-$ decays}},\ }\href
  {https://doi.org/10.1103/PhysRevLett.115.161802} {\bibfield  {journal}
  {\bibinfo  {journal} {Phys. Rev. Lett.}\ }\textbf {\bibinfo {volume} {115}},\
  \bibinfo {pages} {161802} (\bibinfo {year} {2015})},\ \Eprint
  {https://arxiv.org/abs/1508.04094} {arXiv:1508.04094 [hep-ex]} \BibitemShut
  {NoStop}%
\bibitem [{\citenamefont {Cortina~Gil}\ \emph {et~al.}(2021)\citenamefont
  {Cortina~Gil} \emph {et~al.}}]{NA62:2021zjw}%
  \BibitemOpen
  \bibfield  {author} {\bibinfo {author} {\bibfnamefont {E.}~\bibnamefont
  {Cortina~Gil}} \emph {et~al.} (\bibinfo {collaboration} {NA62}),\ }\bibfield
  {title} {\bibinfo {title} {{Measurement of the very rare
  K$^{+}$\textrightarrow{}$ {\pi}^{+}\nu \overline{\nu} $ decay}},\ }\href
  {https://doi.org/10.1007/JHEP06(2021)093} {\bibfield  {journal} {\bibinfo
  {journal} {JHEP}\ }\textbf {\bibinfo {volume} {06}},\ \bibinfo {pages}
  {093}},\ \Eprint {https://arxiv.org/abs/2103.15389} {arXiv:2103.15389
  [hep-ex]} \BibitemShut {NoStop}%
\bibitem [{\citenamefont {Artamonov}\ \emph {et~al.}(2008)\citenamefont
  {Artamonov} \emph {et~al.}}]{E949:2008btt}%
  \BibitemOpen
  \bibfield  {author} {\bibinfo {author} {\bibfnamefont {A.~V.}\ \bibnamefont
  {Artamonov}} \emph {et~al.} (\bibinfo {collaboration} {E949}),\ }\bibfield
  {title} {\bibinfo {title} {{New measurement of the $K^{+} \to \pi^{+} \nu
  \bar{\nu}$ branching ratio}},\ }\href
  {https://doi.org/10.1103/PhysRevLett.101.191802} {\bibfield  {journal}
  {\bibinfo  {journal} {Phys. Rev. Lett.}\ }\textbf {\bibinfo {volume} {101}},\
  \bibinfo {pages} {191802} (\bibinfo {year} {2008})},\ \Eprint
  {https://arxiv.org/abs/0808.2459} {arXiv:0808.2459 [hep-ex]} \BibitemShut
  {NoStop}%
\bibitem [{\citenamefont {Cortina~Gil}\ \emph {et~al.}(2022)\citenamefont
  {Cortina~Gil} \emph {et~al.}}]{NA62:2022qes}%
  \BibitemOpen
  \bibfield  {author} {\bibinfo {author} {\bibfnamefont {E.}~\bibnamefont
  {Cortina~Gil}} \emph {et~al.} (\bibinfo {collaboration} {NA62}),\ }\bibfield
  {title} {\bibinfo {title} {{A measurement of the K$^{+}$\textrightarrow{}
  \ensuremath{\pi}$^{+}$\ensuremath{\mu}$^{+}$\ensuremath{\mu}$^{-}$ decay}},\
  }\href {https://doi.org/10.1007/JHEP06(2023)040} {\bibfield  {journal}
  {\bibinfo  {journal} {JHEP}\ }\textbf {\bibinfo {volume} {11}},\ \bibinfo
  {pages} {011}},\ \bibinfo {note} {[Addendum: JHEP 06, 040 (2023)]},\ \Eprint
  {https://arxiv.org/abs/2209.05076} {arXiv:2209.05076 [hep-ex]} \BibitemShut
  {NoStop}%
\bibitem [{\citenamefont {Alavi-Harati}\ \emph {et~al.}(2004)\citenamefont
  {Alavi-Harati} \emph {et~al.}}]{KTeV:2003sls}%
  \BibitemOpen
  \bibfield  {author} {\bibinfo {author} {\bibfnamefont {A.}~\bibnamefont
  {Alavi-Harati}} \emph {et~al.} (\bibinfo {collaboration} {KTeV}),\ }\bibfield
   {title} {\bibinfo {title} {{Search for the rare decay K(L) ---\ensuremath{>}
  pi0 e+ e-}},\ }\href {https://doi.org/10.1103/PhysRevLett.93.021805}
  {\bibfield  {journal} {\bibinfo  {journal} {Phys. Rev. Lett.}\ }\textbf
  {\bibinfo {volume} {93}},\ \bibinfo {pages} {021805} (\bibinfo {year}
  {2004})},\ \Eprint {https://arxiv.org/abs/hep-ex/0309072}
  {arXiv:hep-ex/0309072} \BibitemShut {NoStop}%
\bibitem [{\citenamefont {Alavi-Harati}\ \emph {et~al.}(2000)\citenamefont
  {Alavi-Harati} \emph {et~al.}}]{KTEV:2000ngj}%
  \BibitemOpen
  \bibfield  {author} {\bibinfo {author} {\bibfnamefont {A.}~\bibnamefont
  {Alavi-Harati}} \emph {et~al.} (\bibinfo {collaboration} {KTEV}),\ }\bibfield
   {title} {\bibinfo {title} {{Search for the Decay $K_L \to \pi^0 \mu^+
  \mu^-$}},\ }\href {https://doi.org/10.1103/PhysRevLett.84.5279} {\bibfield
  {journal} {\bibinfo  {journal} {Phys. Rev. Lett.}\ }\textbf {\bibinfo
  {volume} {84}},\ \bibinfo {pages} {5279} (\bibinfo {year} {2000})},\ \Eprint
  {https://arxiv.org/abs/hep-ex/0001006} {arXiv:hep-ex/0001006} \BibitemShut
  {NoStop}%
\bibitem [{\citenamefont {Merkel}\ \emph {et~al.}(2011)\citenamefont {Merkel}
  \emph {et~al.}}]{A1:2011yso}%
  \BibitemOpen
  \bibfield  {author} {\bibinfo {author} {\bibfnamefont {H.}~\bibnamefont
  {Merkel}} \emph {et~al.} (\bibinfo {collaboration} {A1}),\ }\bibfield
  {title} {\bibinfo {title} {{Search for Light Gauge Bosons of the Dark Sector
  at the Mainz Microtron}},\ }\href
  {https://doi.org/10.1103/PhysRevLett.106.251802} {\bibfield  {journal}
  {\bibinfo  {journal} {Phys. Rev. Lett.}\ }\textbf {\bibinfo {volume} {106}},\
  \bibinfo {pages} {251802} (\bibinfo {year} {2011})},\ \Eprint
  {https://arxiv.org/abs/1101.4091} {arXiv:1101.4091 [nucl-ex]} \BibitemShut
  {NoStop}%
\bibitem [{\citenamefont {Merkel}\ \emph {et~al.}(2014)\citenamefont {Merkel}
  \emph {et~al.}}]{Merkel:2014avp}%
  \BibitemOpen
  \bibfield  {author} {\bibinfo {author} {\bibfnamefont {H.}~\bibnamefont
  {Merkel}} \emph {et~al.},\ }\bibfield  {title} {\bibinfo {title} {{Search at
  the Mainz Microtron for Light Massive Gauge Bosons Relevant for the Muon g-2
  Anomaly}},\ }\href {https://doi.org/10.1103/PhysRevLett.112.221802}
  {\bibfield  {journal} {\bibinfo  {journal} {Phys. Rev. Lett.}\ }\textbf
  {\bibinfo {volume} {112}},\ \bibinfo {pages} {221802} (\bibinfo {year}
  {2014})},\ \Eprint {https://arxiv.org/abs/1404.5502} {arXiv:1404.5502
  [hep-ex]} \BibitemShut {NoStop}%
\bibitem [{\citenamefont {Bodas}\ \emph {et~al.}(2021)\citenamefont {Bodas},
  \citenamefont {Coy},\ and\ \citenamefont {King}}]{Bodas:2021fsy}%
  \BibitemOpen
  \bibfield  {author} {\bibinfo {author} {\bibfnamefont {A.}~\bibnamefont
  {Bodas}}, \bibinfo {author} {\bibfnamefont {R.}~\bibnamefont {Coy}},\ and\
  \bibinfo {author} {\bibfnamefont {S.~J.~D.}\ \bibnamefont {King}},\
  }\bibfield  {title} {\bibinfo {title} {{Solving the electron and muon $g-2$
  anomalies in $Z'$ models}},\ }\href
  {https://doi.org/10.1140/epjc/s10052-021-09850-x} {\bibfield  {journal}
  {\bibinfo  {journal} {Eur. Phys. J. C}\ }\textbf {\bibinfo {volume} {81}},\
  \bibinfo {pages} {1065} (\bibinfo {year} {2021})},\ \Eprint
  {https://arxiv.org/abs/2102.07781} {arXiv:2102.07781 [hep-ph]} \BibitemShut
  {NoStop}%
\bibitem [{\citenamefont {Abrahamyan}\ \emph {et~al.}(2011)\citenamefont
  {Abrahamyan} \emph {et~al.}}]{Abrahamyan:2011gv}%
  \BibitemOpen
  \bibfield  {author} {\bibinfo {author} {\bibfnamefont {S.}~\bibnamefont
  {Abrahamyan}} \emph {et~al.} (\bibinfo {collaboration} {APEX}),\ }\bibfield
  {title} {\bibinfo {title} {{Search for a New Gauge Boson in Electron-Nucleus
  Fixed-Target Scattering by the APEX Experiment}},\ }\href
  {https://doi.org/10.1103/PhysRevLett.107.191804} {\bibfield  {journal}
  {\bibinfo  {journal} {Phys. Rev. Lett.}\ }\textbf {\bibinfo {volume} {107}},\
  \bibinfo {pages} {191804} (\bibinfo {year} {2011})},\ \Eprint
  {https://arxiv.org/abs/1108.2750} {arXiv:1108.2750 [hep-ex]} \BibitemShut
  {NoStop}%
\bibitem [{\citenamefont {Ablikim}\ \emph {et~al.}(2017)\citenamefont {Ablikim}
  \emph {et~al.}}]{Ablikim:2017aab}%
  \BibitemOpen
  \bibfield  {author} {\bibinfo {author} {\bibfnamefont {M.}~\bibnamefont
  {Ablikim}} \emph {et~al.} (\bibinfo {collaboration} {BESIII}),\ }\bibfield
  {title} {\bibinfo {title} {{Dark Photon Search in the Mass Range Between 1.5
  and 3.4 GeV/$c^2$}},\ }\href {https://doi.org/10.1016/j.physletb.2017.09.067}
  {\bibfield  {journal} {\bibinfo  {journal} {Phys. Lett.}\ }\textbf {\bibinfo
  {volume} {B774}},\ \bibinfo {pages} {252} (\bibinfo {year} {2017})},\ \Eprint
  {https://arxiv.org/abs/1705.04265} {arXiv:1705.04265 [hep-ex]} \BibitemShut
  {NoStop}%
\bibitem [{\citenamefont {Lees}\ \emph {et~al.}(2014)\citenamefont {Lees} \emph
  {et~al.}}]{Lees:2014xha}%
  \BibitemOpen
  \bibfield  {author} {\bibinfo {author} {\bibfnamefont {J.~P.}\ \bibnamefont
  {Lees}} \emph {et~al.} (\bibinfo {collaboration} {BaBar}),\ }\bibfield
  {title} {\bibinfo {title} {{Search for a Dark Photon in $e^+e^-$ Collisions
  at BaBar}},\ }\href {https://doi.org/10.1103/PhysRevLett.113.201801}
  {\bibfield  {journal} {\bibinfo  {journal} {Phys. Rev. Lett.}\ }\textbf
  {\bibinfo {volume} {113}},\ \bibinfo {pages} {201801} (\bibinfo {year}
  {2014})},\ \Eprint {https://arxiv.org/abs/1406.2980} {arXiv:1406.2980
  [hep-ex]} \BibitemShut {NoStop}%
\bibitem [{\citenamefont {Lees}\ \emph {et~al.}(2016)\citenamefont {Lees} \emph
  {et~al.}}]{TheBABAR:2016rlg}%
  \BibitemOpen
  \bibfield  {author} {\bibinfo {author} {\bibfnamefont {J.~P.}\ \bibnamefont
  {Lees}} \emph {et~al.} (\bibinfo {collaboration} {BaBar}),\ }\bibfield
  {title} {\bibinfo {title} {{Search for a muonic dark force at BABAR}},\
  }\href {https://doi.org/10.1103/PhysRevD.94.011102} {\bibfield  {journal}
  {\bibinfo  {journal} {Phys. Rev. D}\ }\textbf {\bibinfo {volume} {94}},\
  \bibinfo {pages} {011102} (\bibinfo {year} {2016})},\ \Eprint
  {https://arxiv.org/abs/1606.03501} {arXiv:1606.03501 [hep-ex]} \BibitemShut
  {NoStop}%
\bibitem [{\citenamefont {Bauer}\ \emph {et~al.}(2018)\citenamefont {Bauer},
  \citenamefont {Foldenauer},\ and\ \citenamefont {Jaeckel}}]{Bauer:2018onh}%
  \BibitemOpen
  \bibfield  {author} {\bibinfo {author} {\bibfnamefont {M.}~\bibnamefont
  {Bauer}}, \bibinfo {author} {\bibfnamefont {P.}~\bibnamefont {Foldenauer}},\
  and\ \bibinfo {author} {\bibfnamefont {J.}~\bibnamefont {Jaeckel}},\
  }\bibfield  {title} {\bibinfo {title} {{Hunting All the Hidden Photons}},\
  }\href {https://doi.org/10.1007/JHEP07(2018)094} {\bibfield  {journal}
  {\bibinfo  {journal} {JHEP}\ }\textbf {\bibinfo {volume} {07}},\ \bibinfo
  {pages} {094}},\ \Eprint {https://arxiv.org/abs/1803.05466} {arXiv:1803.05466
  [hep-ph]} \BibitemShut {NoStop}%
\bibitem [{\citenamefont {Bergsma}\ \emph {et~al.}(1985)\citenamefont {Bergsma}
  \emph {et~al.}}]{Bergsma:1985qz}%
  \BibitemOpen
  \bibfield  {author} {\bibinfo {author} {\bibfnamefont {F.}~\bibnamefont
  {Bergsma}} \emph {et~al.} (\bibinfo {collaboration} {CHARM}),\ }\bibfield
  {title} {\bibinfo {title} {{Search for Axion Like Particle Production in
  400-{GeV} Proton - Copper Interactions}},\ }\href
  {https://doi.org/10.1016/0370-2693(85)90400-9} {\bibfield  {journal}
  {\bibinfo  {journal} {Phys. Lett.}\ }\textbf {\bibinfo {volume} {157B}},\
  \bibinfo {pages} {458} (\bibinfo {year} {1985})}\BibitemShut {NoStop}%
\bibitem [{\citenamefont {Tsai}\ \emph {et~al.}(2021)\citenamefont {Tsai},
  \citenamefont {deNiverville},\ and\ \citenamefont {Liu}}]{Tsai:2019buq}%
  \BibitemOpen
  \bibfield  {author} {\bibinfo {author} {\bibfnamefont {Y.-D.}\ \bibnamefont
  {Tsai}}, \bibinfo {author} {\bibfnamefont {P.}~\bibnamefont {deNiverville}},\
  and\ \bibinfo {author} {\bibfnamefont {M.~X.}\ \bibnamefont {Liu}},\
  }\bibfield  {title} {\bibinfo {title} {{Dark Photon and Muon $g-2$ Inspired
  Inelastic Dark Matter Models at the High-Energy Intensity Frontier}},\ }\href
  {https://doi.org/10.1103/PhysRevLett.126.181801} {\bibfield  {journal}
  {\bibinfo  {journal} {Phys. Rev. Lett.}\ }\textbf {\bibinfo {volume} {126}},\
  \bibinfo {pages} {181801} (\bibinfo {year} {2021})},\ \Eprint
  {https://arxiv.org/abs/1908.07525} {arXiv:1908.07525 [hep-ph]} \BibitemShut
  {NoStop}%
\bibitem [{CMS(2019)}]{CMS:2019foo}%
  \BibitemOpen
  \bibfield  {title} {\bibinfo {title} {{Search for a narrow resonance decaying
  to a pair of muons in proton-proton collisions at 13 TeV}},\ }\href@noop {}
  {\  (\bibinfo {year} {2019})}\BibitemShut {NoStop}%
\bibitem [{\citenamefont {Andreas}\ \emph {et~al.}(2012)\citenamefont
  {Andreas}, \citenamefont {Niebuhr},\ and\ \citenamefont
  {Ringwald}}]{Andreas:2012mt}%
  \BibitemOpen
  \bibfield  {author} {\bibinfo {author} {\bibfnamefont {S.}~\bibnamefont
  {Andreas}}, \bibinfo {author} {\bibfnamefont {C.}~\bibnamefont {Niebuhr}},\
  and\ \bibinfo {author} {\bibfnamefont {A.}~\bibnamefont {Ringwald}},\
  }\bibfield  {title} {\bibinfo {title} {{New Limits on Hidden Photons from
  Past Electron Beam Dumps}},\ }\href
  {https://doi.org/10.1103/PhysRevD.86.095019} {\bibfield  {journal} {\bibinfo
  {journal} {Phys. Rev. D}\ }\textbf {\bibinfo {volume} {86}},\ \bibinfo
  {pages} {095019} (\bibinfo {year} {2012})},\ \Eprint
  {https://arxiv.org/abs/1209.6083} {arXiv:1209.6083 [hep-ph]} \BibitemShut
  {NoStop}%
\bibitem [{\citenamefont {Bjorken}\ \emph {et~al.}(2009)\citenamefont
  {Bjorken}, \citenamefont {Essig}, \citenamefont {Schuster},\ and\
  \citenamefont {Toro}}]{Bjorken:2009mm}%
  \BibitemOpen
  \bibfield  {author} {\bibinfo {author} {\bibfnamefont {J.~D.}\ \bibnamefont
  {Bjorken}}, \bibinfo {author} {\bibfnamefont {R.}~\bibnamefont {Essig}},
  \bibinfo {author} {\bibfnamefont {P.}~\bibnamefont {Schuster}},\ and\
  \bibinfo {author} {\bibfnamefont {N.}~\bibnamefont {Toro}},\ }\bibfield
  {title} {\bibinfo {title} {{New Fixed-Target Experiments to Search for Dark
  Gauge Forces}},\ }\href {https://doi.org/10.1103/PhysRevD.80.075018}
  {\bibfield  {journal} {\bibinfo  {journal} {Phys. Rev. D}\ }\textbf {\bibinfo
  {volume} {80}},\ \bibinfo {pages} {075018} (\bibinfo {year} {2009})},\
  \Eprint {https://arxiv.org/abs/0906.0580} {arXiv:0906.0580 [hep-ph]}
  \BibitemShut {NoStop}%
\bibitem [{\citenamefont {Riordan}\ \emph {et~al.}(1987)\citenamefont {Riordan}
  \emph {et~al.}}]{Riordan:1987aw}%
  \BibitemOpen
  \bibfield  {author} {\bibinfo {author} {\bibfnamefont {E.~M.}\ \bibnamefont
  {Riordan}} \emph {et~al.},\ }\bibfield  {title} {\bibinfo {title} {{A Search
  for Short Lived Axions in an Electron Beam Dump Experiment}},\ }\href
  {https://doi.org/10.1103/PhysRevLett.59.755} {\bibfield  {journal} {\bibinfo
  {journal} {Phys. Rev. Lett.}\ }\textbf {\bibinfo {volume} {59}},\ \bibinfo
  {pages} {755} (\bibinfo {year} {1987})}\BibitemShut {NoStop}%
\bibitem [{\citenamefont {Bross}\ \emph {et~al.}(1991)\citenamefont {Bross},
  \citenamefont {Crisler}, \citenamefont {Pordes}, \citenamefont {Volk},
  \citenamefont {Errede},\ and\ \citenamefont {Wrbanek}}]{Bross:1989mp}%
  \BibitemOpen
  \bibfield  {author} {\bibinfo {author} {\bibfnamefont {A.}~\bibnamefont
  {Bross}}, \bibinfo {author} {\bibfnamefont {M.}~\bibnamefont {Crisler}},
  \bibinfo {author} {\bibfnamefont {S.~H.}\ \bibnamefont {Pordes}}, \bibinfo
  {author} {\bibfnamefont {J.}~\bibnamefont {Volk}}, \bibinfo {author}
  {\bibfnamefont {S.}~\bibnamefont {Errede}},\ and\ \bibinfo {author}
  {\bibfnamefont {J.}~\bibnamefont {Wrbanek}},\ }\bibfield  {title} {\bibinfo
  {title} {{A Search for Short-lived Particles Produced in an Electron Beam
  Dump}},\ }\href {https://doi.org/10.1103/PhysRevLett.67.2942} {\bibfield
  {journal} {\bibinfo  {journal} {Phys. Rev. Lett.}\ }\textbf {\bibinfo
  {volume} {67}},\ \bibinfo {pages} {2942} (\bibinfo {year}
  {1991})}\BibitemShut {NoStop}%
\bibitem [{\citenamefont {Petersen}(2023)}]{Petersen:2023hgm}%
  \BibitemOpen
  \bibfield  {author} {\bibinfo {author} {\bibfnamefont {B.}~\bibnamefont
  {Petersen}} (\bibinfo {collaboration} {FASER}),\ }\bibfield  {title}
  {\bibinfo {title} {{First Physics Results from the FASER Experiment}},\ }in\
  \href@noop {} {\emph {\bibinfo {booktitle} {{57th Rencontres de Moriond on
  Electroweak Interactions and Unified Theories}}}}\ (\bibinfo {year} {2023})\
  \Eprint {https://arxiv.org/abs/2305.08665} {arXiv:2305.08665 [hep-ex]}
  \BibitemShut {NoStop}%
\bibitem [{\citenamefont {Abreu}\ \emph {et~al.}(2023)\citenamefont {Abreu}
  \emph {et~al.}}]{FASER:2023tle}%
  \BibitemOpen
  \bibfield  {author} {\bibinfo {author} {\bibfnamefont {H.}~\bibnamefont
  {Abreu}} \emph {et~al.} (\bibinfo {collaboration} {FASER}),\ }\bibfield
  {title} {\bibinfo {title} {{Search for Dark Photons with the FASER detector
  at the LHC}},\ }\href@noop {} {\  (\bibinfo {year} {2023})},\ \Eprint
  {https://arxiv.org/abs/2308.05587} {arXiv:2308.05587 [hep-ex]} \BibitemShut
  {NoStop}%
\bibitem [{\citenamefont {Adrian}\ \emph {et~al.}(2018)\citenamefont {Adrian}
  \emph {et~al.}}]{Adrian:2018scb}%
  \BibitemOpen
  \bibfield  {author} {\bibinfo {author} {\bibfnamefont {P.~H.}\ \bibnamefont
  {Adrian}} \emph {et~al.},\ }\bibfield  {title} {\bibinfo {title} {{Search for
  a Dark Photon in Electro-Produced $e^{+}e^{-}$ Pairs with the Heavy Photon
  Search Experiment at JLab}},\ }\href@noop {} {\  (\bibinfo {year} {2018})},\
  \Eprint {https://arxiv.org/abs/1807.11530} {arXiv:1807.11530 [hep-ex]}
  \BibitemShut {NoStop}%
\bibitem [{\citenamefont {Konaka}\ \emph {et~al.}(1986)\citenamefont {Konaka}
  \emph {et~al.}}]{Konaka:1986cb}%
  \BibitemOpen
  \bibfield  {author} {\bibinfo {author} {\bibfnamefont {A.}~\bibnamefont
  {Konaka}} \emph {et~al.},\ }\bibfield  {title} {\bibinfo {title} {{Search for
  Neutral Particles in Electron Beam Dump Experiment}},\ }\bibfield
  {booktitle} {\emph {\bibinfo {booktitle} {{Proceedings, 23RD International
  Conference on High Energy Physics, JULY 16-23, 1986, Berkeley, CA}}},\ }\href
  {https://doi.org/10.1103/PhysRevLett.57.659} {\bibfield  {journal} {\bibinfo
  {journal} {Phys. Rev. Lett.}\ }\textbf {\bibinfo {volume} {57}},\ \bibinfo
  {pages} {659} (\bibinfo {year} {1986})}\BibitemShut {NoStop}%
\bibitem [{\citenamefont {Anastasi}\ \emph {et~al.}(2015)\citenamefont
  {Anastasi} \emph {et~al.}}]{Anastasi:2015qla}%
  \BibitemOpen
  \bibfield  {author} {\bibinfo {author} {\bibfnamefont {A.}~\bibnamefont
  {Anastasi}} \emph {et~al.},\ }\bibfield  {title} {\bibinfo {title} {{Limit on
  the production of a low-mass vector boson in $\mathrm{e}^{+}\mathrm{e}^{-} o
  \mathrm{U}\gamma$, $\mathrm{U} o \mathrm{e}^{+}\mathrm{e}^{-}$ with the KLOE
  experiment}},\ }\href {https://doi.org/10.1016/j.physletb.2015.10.003}
  {\bibfield  {journal} {\bibinfo  {journal} {Phys. Lett.}\ }\textbf {\bibinfo
  {volume} {B750}},\ \bibinfo {pages} {633} (\bibinfo {year} {2015})},\ \Eprint
  {https://arxiv.org/abs/1509.00740} {arXiv:1509.00740 [hep-ex]} \BibitemShut
  {NoStop}%
\bibitem [{\citenamefont {Anastasi}\ \emph {et~al.}(2016)\citenamefont
  {Anastasi} \emph {et~al.}}]{Anastasi:2016ktq}%
  \BibitemOpen
  \bibfield  {author} {\bibinfo {author} {\bibfnamefont {A.}~\bibnamefont
  {Anastasi}} \emph {et~al.} (\bibinfo {collaboration} {KLOE-2}),\ }\bibfield
  {title} {\bibinfo {title} {{Limit on the production of a new vector boson in
  ightarrowa$, U$ $\mathrm{e^+ e^-} \pi^+\pi^-$ with the KLOE experiment}},\
  }\href {https://doi.org/10.1016/j.physletb.2016.04.019} {\bibfield  {journal}
  {\bibinfo  {journal} {Phys. Lett.}\ }\textbf {\bibinfo {volume} {B757}},\
  \bibinfo {pages} {356} (\bibinfo {year} {2016})},\ \Eprint
  {https://arxiv.org/abs/1603.06086} {arXiv:1603.06086 [hep-ex]} \BibitemShut
  {NoStop}%
\bibitem [{\citenamefont {Anastasi}\ \emph {et~al.}(2018)\citenamefont
  {Anastasi} \emph {et~al.}}]{Anastasi:2018azp}%
  \BibitemOpen
  \bibfield  {author} {\bibinfo {author} {\bibfnamefont {A.}~\bibnamefont
  {Anastasi}} \emph {et~al.} (\bibinfo {collaboration} {KLOE-2}),\ }\bibfield
  {title} {\bibinfo {title} {{Combined limit on the production of a light gauge
  boson decaying into $\mu^+\mu^-$ and $\pi^+\pi^-$}},\ }\href@noop {}
  {\bibfield  {journal} {\bibinfo  {journal} {Submitted to: Phys. Lett. B}\ }
  (\bibinfo {year} {2018})},\ \Eprint {https://arxiv.org/abs/1807.02691}
  {arXiv:1807.02691 [hep-ex]} \BibitemShut {NoStop}%
\bibitem [{\citenamefont {Babusci}\ \emph {et~al.}(2013)\citenamefont {Babusci}
  \emph {et~al.}}]{Babusci:2012cr}%
  \BibitemOpen
  \bibfield  {author} {\bibinfo {author} {\bibfnamefont {D.}~\bibnamefont
  {Babusci}} \emph {et~al.} (\bibinfo {collaboration} {KLOE-2}),\ }\bibfield
  {title} {\bibinfo {title} {{Limit on the production of a light vector gauge
  boson in phi meson decays with the KLOE detector}},\ }\href
  {https://doi.org/10.1016/j.physletb.2013.01.067} {\bibfield  {journal}
  {\bibinfo  {journal} {Phys. Lett.}\ }\textbf {\bibinfo {volume} {B720}},\
  \bibinfo {pages} {111} (\bibinfo {year} {2013})},\ \Eprint
  {https://arxiv.org/abs/1210.3927} {arXiv:1210.3927 [hep-ex]} \BibitemShut
  {NoStop}%
\bibitem [{\citenamefont {Babusci}\ \emph {et~al.}(2014)\citenamefont {Babusci}
  \emph {et~al.}}]{Babusci:2014sta}%
  \BibitemOpen
  \bibfield  {author} {\bibinfo {author} {\bibfnamefont {D.}~\bibnamefont
  {Babusci}} \emph {et~al.} (\bibinfo {collaboration} {KLOE-2}),\ }\bibfield
  {title} {\bibinfo {title} {{Search for light vector boson production in
  $e^+e^- ightarrow \mu^+ \mu^- \gamma$ interactions with the KLOE
  experiment}},\ }\href {https://doi.org/10.1016/j.physletb.2014.08.005}
  {\bibfield  {journal} {\bibinfo  {journal} {Phys. Lett.}\ }\textbf {\bibinfo
  {volume} {B736}},\ \bibinfo {pages} {459} (\bibinfo {year} {2014})},\ \Eprint
  {https://arxiv.org/abs/1404.7772} {arXiv:1404.7772 [hep-ex]} \BibitemShut
  {NoStop}%
\bibitem [{\citenamefont {Aaij}\ \emph {et~al.}(2018)\citenamefont {Aaij} \emph
  {et~al.}}]{Aaij:2017rft}%
  \BibitemOpen
  \bibfield  {author} {\bibinfo {author} {\bibfnamefont {R.}~\bibnamefont
  {Aaij}} \emph {et~al.} (\bibinfo {collaboration} {LHCb}),\ }\bibfield
  {title} {\bibinfo {title} {{Search for Dark Photons Produced in 13 TeV $pp$
  Collisions}},\ }\href {https://doi.org/10.1103/PhysRevLett.120.061801}
  {\bibfield  {journal} {\bibinfo  {journal} {Phys. Rev. Lett.}\ }\textbf
  {\bibinfo {volume} {120}},\ \bibinfo {pages} {061801} (\bibinfo {year}
  {2018})},\ \Eprint {https://arxiv.org/abs/1710.02867} {arXiv:1710.02867
  [hep-ex]} \BibitemShut {NoStop}%
\bibitem [{\citenamefont {Aaij}\ \emph {et~al.}(2019)\citenamefont {Aaij} \emph
  {et~al.}}]{Aaij:2019bvg}%
  \BibitemOpen
  \bibfield  {author} {\bibinfo {author} {\bibfnamefont {R.}~\bibnamefont
  {Aaij}} \emph {et~al.} (\bibinfo {collaboration} {LHCb}),\ }\bibfield
  {title} {\bibinfo {title} {{Search for $A'\! o\!\mu^+\mu^-$ decays}},\
  }\href@noop {} {\  (\bibinfo {year} {2019})},\ \Eprint
  {https://arxiv.org/abs/1910.06926} {arXiv:1910.06926 [hep-ex]} \BibitemShut
  {NoStop}%
\bibitem [{\citenamefont {Batley}\ \emph {et~al.}(2015)\citenamefont {Batley}
  \emph {et~al.}}]{Batley:2015lha}%
  \BibitemOpen
  \bibfield  {author} {\bibinfo {author} {\bibfnamefont {J.~R.}\ \bibnamefont
  {Batley}} \emph {et~al.} (\bibinfo {collaboration} {NA48/2}),\ }\bibfield
  {title} {\bibinfo {title} {{Search for the dark photon in $\pi^0$ decays}},\
  }\href {https://doi.org/10.1016/j.physletb.2015.04.068} {\bibfield  {journal}
  {\bibinfo  {journal} {Phys. Lett.}\ }\textbf {\bibinfo {volume} {B746}},\
  \bibinfo {pages} {178} (\bibinfo {year} {2015})},\ \Eprint
  {https://arxiv.org/abs/1504.00607} {arXiv:1504.00607 [hep-ex]} \BibitemShut
  {NoStop}%
\bibitem [{\citenamefont {D"obrich}\ \emph {et~al.}(2023)\citenamefont
  {D"obrich}, \citenamefont {Minucci},\ and\ \citenamefont
  {Spadaro}}]{Dobrich:2023dkm}%
  \BibitemOpen
  \bibfield  {author} {\bibinfo {author} {\bibfnamefont {B.}~\bibnamefont
  {D"obrich}}, \bibinfo {author} {\bibfnamefont {E.}~\bibnamefont {Minucci}},\
  and\ \bibinfo {author} {\bibfnamefont {T.}~\bibnamefont {Spadaro}} (\bibinfo
  {collaboration} {NA62}),\ }\bibfield  {title} {\bibinfo {title} {{Search for
  dark photon decays to $\mu^+\mu^-$ at NA62}},\ }\href@noop {} {\  (\bibinfo
  {year} {2023})},\ \Eprint {https://arxiv.org/abs/2303.08666}
  {arXiv:2303.08666 [hep-ex]} \BibitemShut {NoStop}%
\bibitem [{\citenamefont {Banerjee}\ \emph {et~al.}(2018)\citenamefont
  {Banerjee} \emph {et~al.}}]{Banerjee:2018vgk}%
  \BibitemOpen
  \bibfield  {author} {\bibinfo {author} {\bibfnamefont {D.}~\bibnamefont
  {Banerjee}} \emph {et~al.} (\bibinfo {collaboration} {NA64}),\ }\bibfield
  {title} {\bibinfo {title} {{Search for a new X(16.7) boson and dark photons
  in the NA64 experiment at CERN}},\ }\href@noop {} {\  (\bibinfo {year}
  {2018})},\ \Eprint {https://arxiv.org/abs/1803.07748} {arXiv:1803.07748
  [hep-ex]} \BibitemShut {NoStop}%
\bibitem [{\citenamefont {Banerjee}\ \emph {et~al.}(2019)\citenamefont
  {Banerjee} \emph {et~al.}}]{Banerjee:2019hmi}%
  \BibitemOpen
  \bibfield  {author} {\bibinfo {author} {\bibfnamefont {D.}~\bibnamefont
  {Banerjee}} \emph {et~al.},\ }\bibfield  {title} {\bibinfo {title} {{Improved
  limits on a hypothetical X(16.7) boson and a dark photon decaying into
  $e^+e^-$ pairs}},\ }\href@noop {} {\  (\bibinfo {year} {2019})},\ \Eprint
  {https://arxiv.org/abs/1912.11389} {arXiv:1912.11389 [hep-ex]} \BibitemShut
  {NoStop}%
\bibitem [{\citenamefont {Astier}\ \emph {et~al.}(2001)\citenamefont {Astier}
  \emph {et~al.}}]{Astier:2001ck}%
  \BibitemOpen
  \bibfield  {author} {\bibinfo {author} {\bibfnamefont {P.}~\bibnamefont
  {Astier}} \emph {et~al.} (\bibinfo {collaboration} {NOMAD}),\ }\bibfield
  {title} {\bibinfo {title} {{Search for heavy neutrinos mixing with tau
  neutrinos}},\ }\href {https://doi.org/10.1016/S0370-2693(01)00362-8}
  {\bibfield  {journal} {\bibinfo  {journal} {Phys. Lett.}\ }\textbf {\bibinfo
  {volume} {B506}},\ \bibinfo {pages} {27} (\bibinfo {year} {2001})},\ \Eprint
  {https://arxiv.org/abs/hep-ex/0101041} {arXiv:hep-ex/0101041 [hep-ex]}
  \BibitemShut {NoStop}%
\bibitem [{\citenamefont {Blumlein}\ \emph {et~al.}(1991)\citenamefont
  {Blumlein} \emph {et~al.}}]{Blumlein:1990ay}%
  \BibitemOpen
  \bibfield  {author} {\bibinfo {author} {\bibfnamefont {J.}~\bibnamefont
  {Blumlein}} \emph {et~al.},\ }\bibfield  {title} {\bibinfo {title} {{Limits
  on neutral light scalar and pseudoscalar particles in a proton beam dump
  experiment}},\ }\href {https://doi.org/10.1007/BF01548556} {\bibfield
  {journal} {\bibinfo  {journal} {Z. Phys.}\ }\textbf {\bibinfo {volume}
  {C51}},\ \bibinfo {pages} {341} (\bibinfo {year} {1991})}\BibitemShut
  {NoStop}%
\bibitem [{\citenamefont {Blumlein}\ \emph {et~al.}(1992)\citenamefont
  {Blumlein} \emph {et~al.}}]{Blumlein:1991xh}%
  \BibitemOpen
  \bibfield  {author} {\bibinfo {author} {\bibfnamefont {J.}~\bibnamefont
  {Blumlein}} \emph {et~al.},\ }\bibfield  {title} {\bibinfo {title} {{Limits
  on the mass of light (pseudo)scalar particles from Bethe-Heitler e+ e- and
  mu+ mu- pair production in a proton - iron beam dump experiment}},\ }\href
  {https://doi.org/10.1142/S0217751X9200171X} {\bibfield  {journal} {\bibinfo
  {journal} {Int. J. Mod. Phys.}\ }\textbf {\bibinfo {volume} {A7}},\ \bibinfo
  {pages} {3835} (\bibinfo {year} {1992})}\BibitemShut {NoStop}%
\bibitem [{\citenamefont {Davier}\ and\ \citenamefont
  {Nguyen~Ngoc}(1989)}]{Davier:1989wz}%
  \BibitemOpen
  \bibfield  {author} {\bibinfo {author} {\bibfnamefont {M.}~\bibnamefont
  {Davier}}\ and\ \bibinfo {author} {\bibfnamefont {H.}~\bibnamefont
  {Nguyen~Ngoc}},\ }\bibfield  {title} {\bibinfo {title} {{An Unambiguous
  Search for a Light Higgs Boson}},\ }\href
  {https://doi.org/10.1016/0370-2693(89)90174-3} {\bibfield  {journal}
  {\bibinfo  {journal} {Phys. Lett.}\ }\textbf {\bibinfo {volume} {B229}},\
  \bibinfo {pages} {150} (\bibinfo {year} {1989})}\BibitemShut {NoStop}%
\bibitem [{\citenamefont {Bernardi}\ \emph {et~al.}(1986)\citenamefont
  {Bernardi} \emph {et~al.}}]{Bernardi:1985ny}%
  \BibitemOpen
  \bibfield  {author} {\bibinfo {author} {\bibfnamefont {G.}~\bibnamefont
  {Bernardi}} \emph {et~al.},\ }\bibfield  {title} {\bibinfo {title} {{Search
  for Neutrino Decay}},\ }\href {https://doi.org/10.1016/0370-2693(86)91602-3}
  {\bibfield  {journal} {\bibinfo  {journal} {Phys. Lett.}\ }\textbf {\bibinfo
  {volume} {166B}},\ \bibinfo {pages} {479} (\bibinfo {year}
  {1986})}\BibitemShut {NoStop}%
\bibitem [{\citenamefont {Ilten}\ \emph {et~al.}(2018)\citenamefont {Ilten},
  \citenamefont {Soreq}, \citenamefont {Williams},\ and\ \citenamefont
  {Xue}}]{Ilten:2018crw}%
  \BibitemOpen
  \bibfield  {author} {\bibinfo {author} {\bibfnamefont {P.}~\bibnamefont
  {Ilten}}, \bibinfo {author} {\bibfnamefont {Y.}~\bibnamefont {Soreq}},
  \bibinfo {author} {\bibfnamefont {M.}~\bibnamefont {Williams}},\ and\
  \bibinfo {author} {\bibfnamefont {W.}~\bibnamefont {Xue}},\ }\bibfield
  {title} {\bibinfo {title} {{Serendipity in dark photon searches}},\ }\href
  {https://doi.org/10.1007/JHEP06(2018)004} {\bibfield  {journal} {\bibinfo
  {journal} {JHEP}\ }\textbf {\bibinfo {volume} {06}},\ \bibinfo {pages}
  {004}},\ \Eprint {https://arxiv.org/abs/1801.04847} {arXiv:1801.04847
  [hep-ph]} \BibitemShut {NoStop}%
\bibitem [{\citenamefont {Okada}\ and\ \citenamefont
  {Seto}(2023)}]{Okada:2023mdv}%
  \BibitemOpen
  \bibfield  {author} {\bibinfo {author} {\bibfnamefont {N.}~\bibnamefont
  {Okada}}\ and\ \bibinfo {author} {\bibfnamefont {O.}~\bibnamefont {Seto}},\
  }\bibfield  {title} {\bibinfo {title} {{Intergenerational gauged $B-L$ model
  and its implication to muon $g-2$ anomaly and thermal dark matter}},\
  }\href@noop {} {\  (\bibinfo {year} {2023})},\ \Eprint
  {https://arxiv.org/abs/2307.14053} {arXiv:2307.14053 [hep-ph]} \BibitemShut
  {NoStop}%
\bibitem [{\citenamefont {Raby}(2022)}]{Raby2022}%
  \BibitemOpen
  \bibfield  {author} {\bibinfo {author} {\bibfnamefont {S.}~\bibnamefont
  {Raby}},\ }\bibfield  {title} {\bibinfo {title} {{Introduction to the
  standard model and beyond: quantum field theory, symmetries and
  phenomenology}},\ }\href {https://doi.org/10.1080/00107514.2022.2038674}
  {\bibfield  {journal} {\bibinfo  {journal} {Contemp. Phys.}\ }\textbf
  {\bibinfo {volume} {62}},\ \bibinfo {pages} {123} (\bibinfo {year}
  {2022})}\BibitemShut {NoStop}%
\bibitem [{\citenamefont {Schwartz}(2014)}]{Schwartz:2014sze}%
  \BibitemOpen
  \bibfield  {author} {\bibinfo {author} {\bibfnamefont {M.~D.}\ \bibnamefont
  {Schwartz}},\ }\href@noop {} {\emph {\bibinfo {title} {{Quantum Field Theory
  and the Standard Model}}}}\ (\bibinfo  {publisher} {Cambridge University
  Press},\ \bibinfo {year} {2014})\BibitemShut {NoStop}%
\end{thebibliography}%

\end{document}